\documentclass{emulateapj}

\usepackage{graphics}

\shorttitle{Modeling the \ion{Ne}{9} Triplet with HETGS, LETGS, and RGS}
\shortauthors{Ness, Brickhouse, Drake, \& Huenemoerder}

\begin{document}

\title{Modeling the \ion{Ne}{9} Triplet Spectral Region of Capella
with the Chandra and XMM-Newton Gratings}

\author{Jan-Uwe Ness}
\affil{Hamburger Sternwarte, Universit\"at Hamburg, Gojenbergsweg 112, D-21029 Hamburg, Germany}
\email{jness@hs.uni-hamburg.de}

\author{Nancy S. Brickhouse and Jeremy J. Drake}
\affil{Harvard-Smithsonian Center for Astrophysics, 60 Garden Street, Cambridge, MA 02138}
\email{[bhouse,jdrake]@head-cfa.harvard.edu}

\author{David P. Huenemoerder}
\affil{MIT Center for Space Research, 70 Vassar Street, Cambridge, MA 02139}
\email{dph@space.mit.edu}

\begin{abstract}

High resolution X-ray spectroscopy with the diffraction gratings of
Chandra and XMM-Newton offers new chances to study a large variety
of stellar coronal phenomena. A popular X-ray calibration target is
Capella, which has been observed with all gratings with significant
exposure times. We gathered together all available data of the HETGS
(155\,ks), LETGS (219\,ks), and RGS (53\,ks) for comparative
analysis focusing on the \ion{Ne}{9} triplet at around 13.5\,\AA, a
region that is severely blended by strong iron lines. We identify 18
emission lines in this region of the HEG spectrum, including many
from \ion{Fe}{19}, and find good agreement with predictions from a
theoretical model constructed using the Astrophysical Plasma
Emission Code (APEC). The model uses an emission measure
distribution derived from \ion{Fe}{15} to \ion{Fe}{24} lines. The
success of the model is due in part to the inclusion of accurate
wavelengths from laboratory measurements.  While these 18 emission
lines cannot be isolated in the LETGS or RGS spectra, their
wavelengths and fluxes as measured with HEG are consistent with the
lower resolution spectra.

In the Capella model for HEG, the weak intercombination line of
\ion{Ne}{9} is significantly blended by iron lines, which
contribute about half the flux. After accounting for blending in the
He-like diagnostic lines, we find the density to be consistent with
the low density limit ($n_e < 2 \times 10^{10}$ cm$^{-3}$); however,
the electron temperature indicated by the \ion{Ne}{9} $G$-ratio is
surprisingly low ($\sim 2$ MK) compared with the peak of the emission
measure distribution ($\sim 6$ MK).  Models show that the \ion{Ne}{9}
triplet is less blended in cooler plasmas and in plasmas with an
enhanced neon-to-iron abundance ratio.

\end{abstract}


\section{Introduction}

Investigation of stellar coronae in the X-ray wavelength band has
until very recently been restricted to instruments with intrinsically
low spectral resolution, such as the proportional counters and CCDs on
satellites such as Einstein, ROSAT, ASCA, and BeppoSAX.\footnote{
Exceptions are the Einstein spectra of Capella with much better
resolution, obtained by \cite{mewe82} with the objective grating
spectrometer (OGS), covering the range 5 to 30 \AA\ with a resolution
$<$1\,\AA, and of \cite{vedder83} with the focal-plane crystal
spectrometer.}  The limitations of low resolution spectra have restricted
X-ray studies of stellar coronae to measurements of luminosities,
plasma temperatures and estimates of elemental abundances.  The
spatial resolution of X-ray emitting plasma that is routinely
available for studies of the solar corona with satellites such as SOHO
and TRACE is still today a dream of X-ray astronomers.

Despite limited spectral information, substantial progress has been
made in understanding the gross characteristics of stellar coronae
throughout the HR diagram. For example, \cite{1984A&A...138..258S}
showed from Einstein IPC observations of a sample of 34 late-type
stars that coronal temperature was directly correlated with X-ray
luminosity and stellar rotation rate. Similar results were obtained
for a larger sample by \citet{schm90}, and later based on the
so-called hardness ratio obtained from ROSAT PSPC observations by
\citet{schm97}.

These results raised a question as to the nature of the high
temperature plasma in active stellar coronae.
\cite{1978ARA&A..16..393V} had pointed out that the Sun completely
covered with active regions would have an X-ray luminosity of
$\sim\,2\times 10^{29}$\,erg\,s$^{-1}$.  However, the most active
solar-like stars can have X-ray luminosities up to two orders of
magnitude higher than this and so cannot simply be scaled-up versions
of the solar corona. The hot plasma on the most active stars must be
structured differently from that in typical solar active regions. The
radiative loss of a hot, optically-thin, collision-dominated plasma is
essentially proportional to the volume emission measure, defined as
the product of electron density squared and the emitting volume,
$n_e^2V$.  Increasing either the emitting volumes or plasma density
results in an increase in X-ray luminosity.

A key plasma parameter for inferring the size of X-ray emitting regions
is therefore the plasma density. Direct spectroscopic information on
plasma densities at coronal temperatures on stars other than the Sun
first became possible with the advent of ``high'' resolution spectra
($\lambda/\Delta\lambda \sim\,200$) obtained by EUVE that were capable
of separating individual spectral lines.  Even with this resolution,
the available diagnostics have often tended to be less than
definitive, owing to the poor signal-to-noise ratio of observed
spectra or to blended lines. This has been especially the case for the
active stars. Studies of density-sensitive lines of \ion{Fe}{19} to
\ion{Fe}{22} in EUVE spectra of RS~CVn stars revealed tempting
evidence for high densities of $n_e\sim\,10^{12}$ to
$10^{13}$\,cm$^{-3}$ at coronal temperatures near $10^7$\,K
\citep{1993ApJ...418L..41D, drake96}.  Such high densities suggest
that emitting structures are compact: static loop models such as those
described by \cite{rtv} would have heights of $\sim\,1000$\,km with
confining field strengths of up to 1\,kG and surface filling factors
of 1 to 10\%. In the case of the evolved active binary Capella,
\cite{1996aeu..conf..105B} obtained $n_e\sim\,10^{12}$\,cm$^{-3}$ from
\ion{Fe}{19} to \ion{Fe}{22}, but $n_e\sim\,10^{9}$\,cm$^{-3}$ based
on \ion{Fe}{12} to \ion{Fe}{14}, suggesting that the cool and hot
plasma are from distinctly different structures. In contrast to the
high densities reported by \cite{1993ApJ...418L..41D} only upper limits
were found by \cite{mewe01} ($n_e < 2-5 \times 10^{12}$ cm$^{-3}$) using the same line ratios of \ion{Fe}{20}
to \ion{Fe}{22} but with the higher spectral resolution of Chandra LETGS
spectra. They point out that the \ion{Fe}{19} to \ion{Fe}{22} line
ratios are only sensitive above $10^{11}$\,cm$^{-3}$, such that no
tracer for low densities for the hotter plasma component is available.

The high spectral resolution of Chandra and XMM-Newton has
significantly changed the situation regarding plasma diagnostics for
stellar coronae.  High spectral resolution coupled with large
effective area allows application of line-based diagnostic techniques
at X-ray wavelengths.  Line ratios of the strong H-like Lyman series,
which are sensitive to the Boltzmann factor, and the He-like triplets,
which exploit the competition between collisional excitation and
recombination-driven cascades, can be useful temperature diagnostics.
Simply examining the presence or absence of principle lines of
different ionic states of different elements gives an indication of
the emitting plasma temperatures.  The He-like systems also provide
density diagnostics based on the low-lying metastable level
1s2s$\,^3$S$_1$ above the 1s$^2\,^1$S$_0$ ground state. The He-like
oxygen diagnostic has been used to support the previous EUVE result
for Capella, showing that the lower temperature plasma ($\sim$2~MK)
also has lower density \citep{2000ApJ...539L..41C, 2000ApJ...530L.111B, ness01,
2001A&A...365L.329A, brickhouse02}.

We compare the capabilities of this current generation of high
spectral resolution X-ray instruments with a detailed study of the
\ion{Ne}{9} triplet as a density and temperature diagnostic. We focus
our study on the Capella binary system (HD\,34029; $\alpha$ Aurigae;
G8 III + G1 III). Capella has been extensively studied from X-ray to
radio wavelengths, and is the brightest, steady coronal source in the
X-ray sky. As such, it has been a key calibration target for the
Chandra high-energy and low-energy transmission grating spectrometers
(HETGS and LETGS, respectively), as well as the XMM-Newton reflection
grating spectrometers (RGS), and has been observed on several
occasions by both satellites \citep{2000ApJ...539L..41C,
2000ApJ...530L.111B, 2001A&A...365L.329A}. The emission measure
distribution of Capella shows a steep but narrow enhancement at 6\,MK
\citep{1993ApJ...418L..41D, 2000ApJ...530..387B}, making it ideal for
studying the blending of \ion{Ne}{9} with high temperature lines.

The purpose of this study is threefold: (i) assess the accuracy and
reliability of state-of-the-art plasma radiative loss models for
describing the emission spectrum of Capella in the region of the
He-like complex of Ne; (ii) determine how well these models might
describe the spectra of stars both more and less active than Capella
and with different coronal temperatures; and (iii) gain further
insight into the plasma density in the Capella coronae in the key
temperature range 4 to 6\,MK.

\section{He-like Ions in High Resolution Stellar Spectra}
\subsection{The Use of the He-like Triplet Diagnostics}
\label{usehe}

He-like ions are produced at temperatures ranging between $\sim2$\,MK
for \ion{C}{5} and \ion{N}{6} and 10\,MK for \ion{Si}{13}. The
densities over which the diagnostics are sensitive increase with
increasing element number Z, such that the lower Z ions \ion{C}{5},
\ion{N}{6}, and \ion{O}{7} with temperatures of peak emissivity at
1.0, 1.6, and 2.0\,MK, respectively, provide diagnostics for densities
up to $\sim 10^{12}$\,cm$^{-3}$. The higher temperature ($> 6$\,MK)
ions \ion{Mg}{11} and \ion{Si}{13} are sensitive at densities above
$\sim 10^{12}$\,cm$^{-3}$. \ion{Ne}{9}, typically formed at $\sim
4$\,MK, is sensitive to densities between $\sim 10^{11}$ and
$10^{13}$\,cm$^{-3}$. In many cases measurements of \ion{Ne}{9} are
the only chance to fill the gap between the cooler plasma and the
hotter plasma. In addition, some stars show a neon overabundance
\citep[e.g.,][]{2001A&A...365L.324B, 2001ApJ...548L..81D,
2001A&A...365L.336G, 2001ApJ...559.1135H}, making \ion{Ne}{9} lines
more easily detectable in those cases. However, blending with Fe lines
in the \ion{Ne}{9} triplet spectral region can compromise the
diagnostic utility when analyzing plasmas hot enough to produce them
(especially \ion{Fe}{19} at $\sim 6$\,MK).

\begin{figure}[!ht]
\resizebox{\hsize}{!}{\rotatebox{90}{\includegraphics{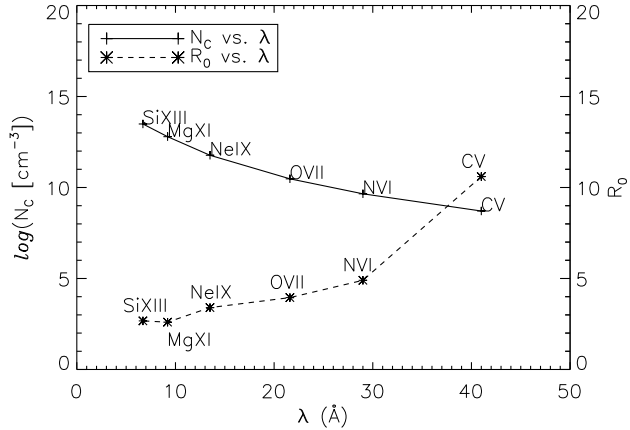}}}
\caption{\label{he_prop}The low-density limits $R_0$
and the critical densities $N_c$ for the different He-like ions show
systematic, anti-correlated trends with wavelength, due ultimately
to the nuclear charge \citep[from][]{ps81,bl72}.  Large $R_0$
(\ion{C}{5}) implies that the $i$ flux may be difficult to measure.
Large $N_c$ may be outside the interesting coronal range
(\ion{Si}{13}).  \ion{Ne}{9} appears to have ideal values for
coronal physics.
}
\end{figure}

The theory of the He-like triplets was originally developed by
\cite{gj69}. The density diagnostic is often referred to as the
$R$-ratio, where $R=f/i$, denoting the forbidden line flux with $f$
and the intercombination line flux with $i$. The relation can be
parameterized as
\begin{equation}
R \ = \frac {R_{\rm 0}} {1 + n_e/N_c}\,,
\end{equation}
with low-density limit $R_{\rm 0}$ and electron density $n_e$. The
critical density $N_c$ is the density at which $R=1/2 R_{\rm 0}$.
\cite{ps81} and \cite{bl72} calculated theoretical values of $R_{\rm
0}$ and $N_c$ and determined the Z-dependence of the
density-sensitivity as shown in Figure~\ref{he_prop}.  Larger
$R$-ratios mean relatively weaker intercombination lines; since the
$i$ is intrinsically weak at low densities, it may be difficult to
measure accurately.

Strong ultraviolet radiation fields can also de-populate the
metastable levels of the He-like triplets and change the $R$-ratios.
For the case of Capella and \ion{Ne}{9}, such fields can be neglected
\citep{ness01}.

For an optically thin plasma, a temperature diagnostic is
\begin{equation}
G\ =\ \frac{i+f}{r}\,,
\end{equation}
where $r$ is the resonance line flux. In collisional ionization
equilibrium (CIE), the $G$-ratio decreases with temperature primarily
because the excitation of the triplet levels (by dielectronic
recombination-driven cascades) decreases faster than the collisional
excitation to the $^1P_1$ \citep[see][]{2001ApJ...556L..91S}. Assuming
{\em a priori} that the plasma is in CIE, the $G$-ratio is a direct
diagnostic of the temperature of \ion{Ne}{9} emission and can be
compared to predictions based on an emission measure distribution.

The He-like triplets have been used for measuring temperatures and
densities in the solar corona. Both the \ion{O}{7} and \ion{Ne}{9}
triplets have been studied for quiescent emission and flares
\citep[e.g.,][]{acton72, keen87}. In solar flares, the \ion{Ne}{9}
lines appeared blended. \citet{mckenz85} and \citet{doyle86} suggested
several candidate lines for these blends.

Generally, stars with lower temperature coronae do not show
significant blending around the \ion{Ne}{9} triplet. Procyon
\citep{ness02b} is a good example. In hotter coronae, the blending is
more problematic. To date, the $R$-ratios measured from MEG spectra of
Capella have not been used to determine densities
\citep{phil01,2001ApJ...549..554A}. \cite{ness02} have attempted to
disentangle the blending for the LETGS/HRC-S spectrum of Algol by
assuming {\it a priori} a $G$-ratio of 0.8.

\subsection{Comparison among the New Instruments}
\label{instr}

\begin{figure}[!ht]
\resizebox{\hsize}{!}{\includegraphics{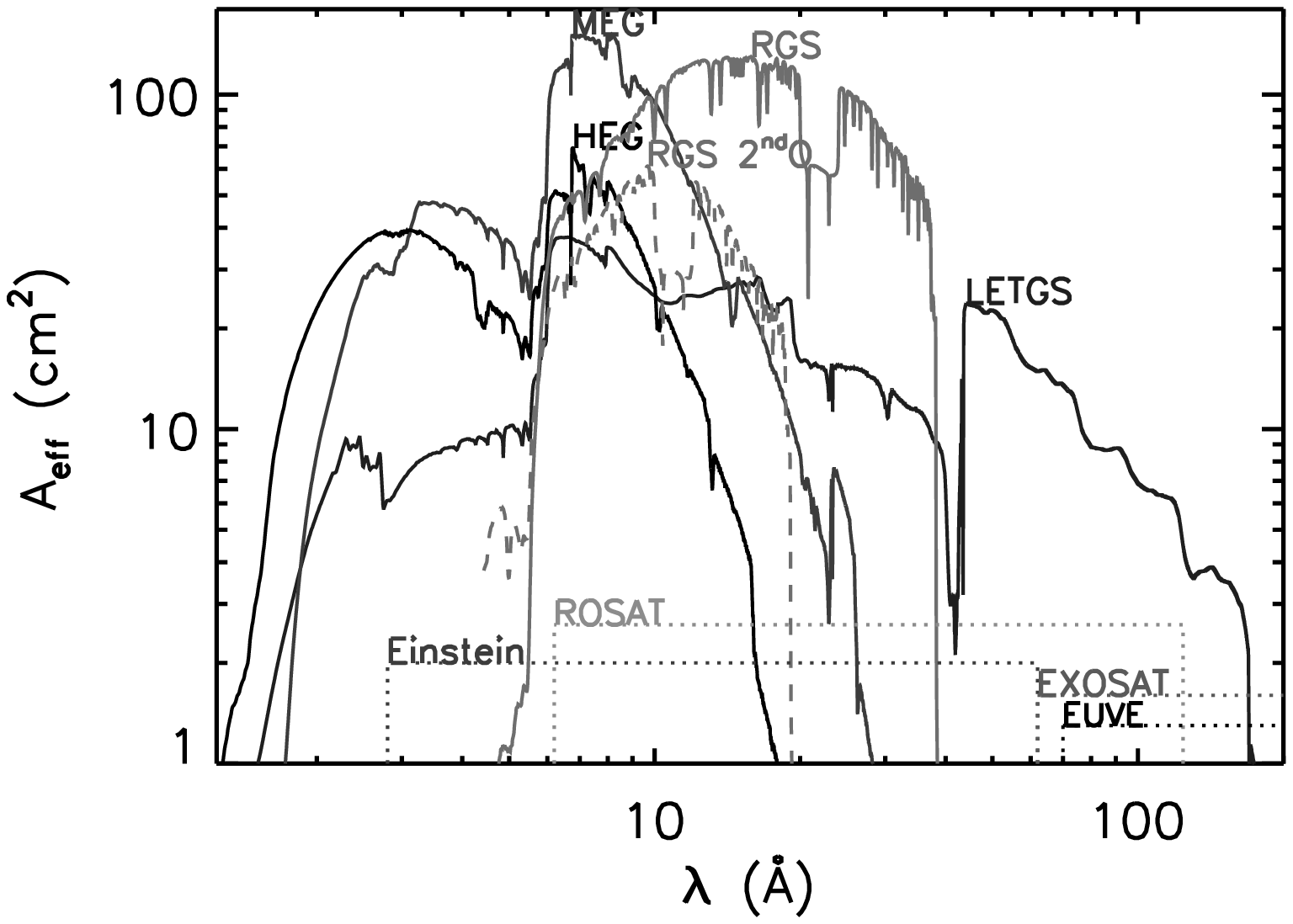}}

\resizebox{\hsize}{!}{\includegraphics{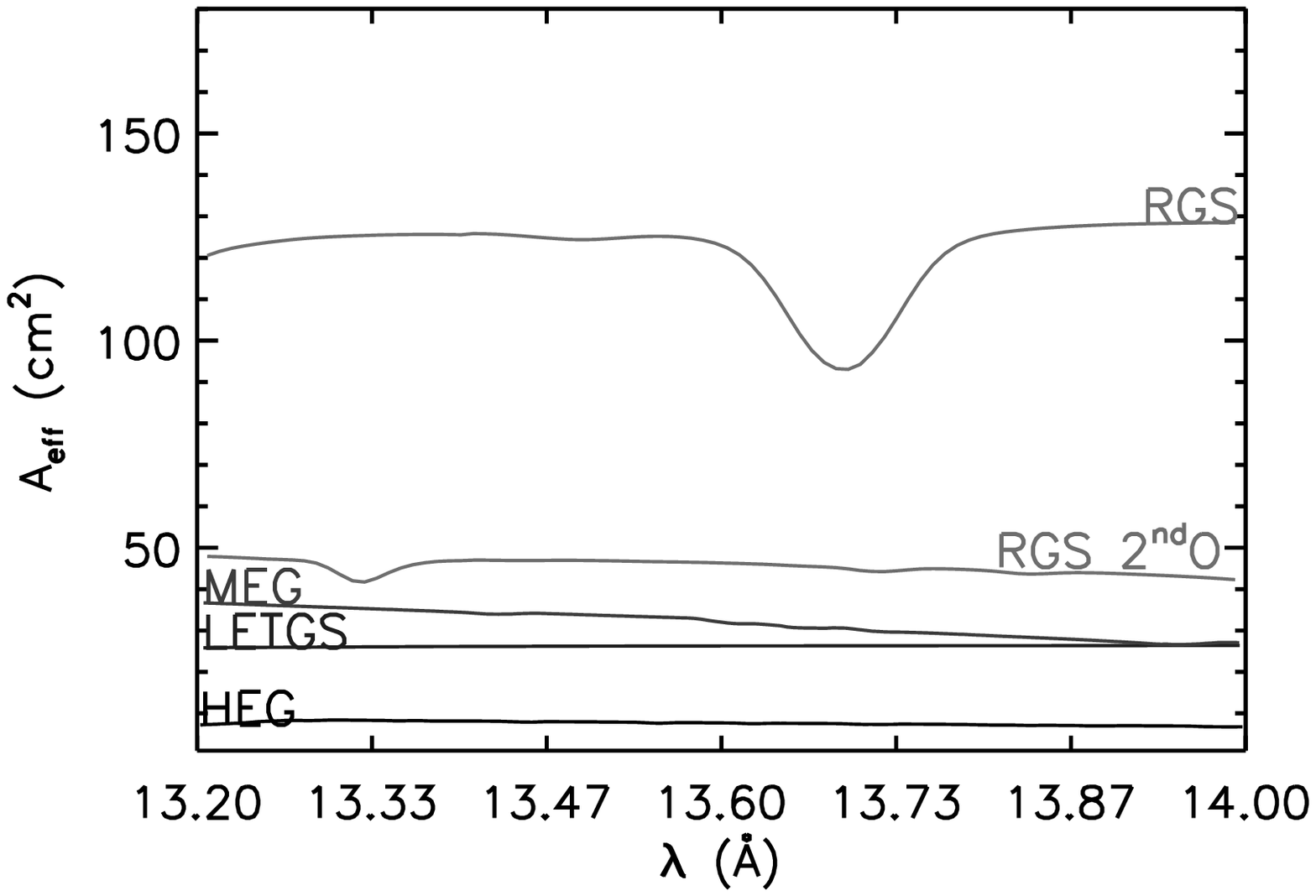}}
\caption{\label{aeff}Effective areas for the Chandra and XMM-Newton gratings,
with ranges of other observatories shown as dotted lines. At 13.6\,\AA, the
RGS has the largest area, while HEG is the least sensitive.  The
bottom panel shows detail in the \ion{Ne}{9} region.}
\end{figure}

The new generation of X-ray telescopes of XMM-Newton and Chandra has
opened new dimensions in sensitivity and resolution.  Spectroscopic
measurements can be carried out with both instruments using the
intrinsic resolution of the CCDs (ACIS, EPIC) and the dispersive
instruments for higher resolution. Three gratings are providing data:
the RGS on board XMM-Newton and the HETGS and LETGS on Chandra. The
HETGS consists of two sets of gratings with different periods, the
High Energy Grating (HEG) and the Medium Energy Grating (MEG), which
intercept X-rays from the inner and outer mirror shells, respectively,
and thus are used concurrently. The RGS and HETGS use the EPIC and
ACIS-S CCD detectors, respectively, while the LETGS can use either
ACIS-S or the micro-channel plate detector HRC-S. Figure~\ref{aeff}
shows the effective areas and the wavelength ranges for the grating
instruments.  While RGS, HEG, and MEG operate in the spectral range
$\la 40\,$\AA, LETGS covers a much larger wavelength range from 2 to
175\,\AA.  All the gratings have uniform resolving power for their
entire wavelength range.

As can be seen from Figure~\ref{aeff} the large wavelength range
covered by LETGS allows the extraction of total fluxes, luminosities,
and hardness ratios corresponding to ROSAT (5 -- 124\,\AA) or Einstein
(3 -- 84\,\AA). With LETGS all He-like triplets ranging from
\ion{C}{5} and \ion{N}{6} up to \ion{Si}{13} can be measured
simultaneously. LETGS also obtains $\Delta n=0$ Fe L-shell lines,
including density-sensitive lines of \ion{Fe}{19} to \ion{Fe}{22},
previously obtained at five times lower spectral resolution with EUVE.

Since LETGS cannot sort orders via detector intrinsic energy
resolution, integrated fluxes and luminosities are still
source-model-dependent.  For a $\log T=6.8$ plasma, the apparent
energy flux obtained by dividing counts by the first order effective
area is about 20\% larger than the true first order flux, when 11
orders are accounted for.  In this ideal isothermal model, only 80\%
of the counts are from first order.

An advantage of the HETGS is its very high spectral resolution near
1\,keV. In addition, order sorting for HEG, MEG, and the RGS is
possible using the energy resolution of the CCD detectors, which is
not possible for the LETGS/HRC-S configuration.  This allows accurate
determination of source-model-independent broadband fluxes and
luminosities.

The RGS covers a wavelength range similar to Einstein with especially
large effective areas above 10\,\AA\ (Fig.~\ref{aeff}). The spectral
resolution of the RGS is comparable to that of LETGS in their region
of overlap.

The mirror point spread function primarily defines the dispersed line
profile of the HEG, MEG, and LEG.  In the case of the RGS, significant
scattering wings also arise from the reflection gratings themselves.
Grating period variance, detector pixelization, and aspect
reconstruction are additional factors.  For \ion{Ne}{9}, the RGS offers the highest
sensitivity, even in second dispersion order; MEG and LETGS have
larger effective areas than HEG. The spectral resolution of the HEG,
however, is the most important factor for a realistic assessment of
the effects of blending. Here we explore the diagnostic utility for
observations of \ion{Ne}{9} with each instrument.

\subsection{Spectral Models}
\label{apec}

Atomic data are fundamental to interpretation of the observed spectra.
We compare observations with models produced using the Astrophysical
Plasma Emission Code \citep{2001ApJ...556L..91S} Version
1.2.\footnote{Available at http://cxc.harvard.edu/atomdb.} APEC
incorporates collisional and radiative rate data appropriate for
modeling optically thin plasmas under the conditions of collisional
ionization equilibrium. The code solves the level-to-level rate matrix
to obtain level populations and produces line emissivities as
functions of electron temperature and density. APEC also calculates
spectral continuum emission from bremsstrahlung, radiative
recombination, and two-photon emission.

We rely on APEC inclusion of the H- and He-like atomic data discussed
by \cite{2001ApJ...556L..91S}, the HULLAC calculations of D.~Liedahl
for the Fe L-shell ions with additional R-matrix calculations
available from CHIANTI V2.0 \citep{1999A&AS..135..171L}, and isosequence scaling
for Ni L-shell ions.  Our models assume the ionization balance of
\citet{1998A&AS..133..403M} and the solar abundance model of
\citet{1989GeCoA..53..197A}. Reference wavelengths are from quantum
electrodynamic calculations for H- and He-like ions
\citep{1988CaJPh..66..586D, ericsson} and from laboratory measurements
for emission lines of Fe L-shell \citep{1998ApJ...502.1015B,
2002ApJS..140..589B} and Ni L-shell ions \citep{2000Shirai}.
Additional L-shell wavelengths are derived from HULLAC energy levels.

Despite significant improvements to spectral modeling over the past
several years, the theoretical atomic data in APEC and other plasma
models remain largely untested over the broad range of applicable
densities and temperatures. The deep observations of three late-type
stars (Capella, Procyon, and HR~1099), used for in-flight calibration
measurements of the dispersion relation and line spread functions of
the gratings, are also useful for determining the extent of agreement
between models and observations and for highlighting issues that
might require additional atomic physics work.  The accuracy of each
rate as well as the completeness of line lists is important to the
correct interpretation of spectral diagnostics, especially at lower
spectral resolution \citep[e.g.,][]{2000ApJ...530..387B}.  This work
is part of a comprehensive effort known as the ``Emission Line
Project'' to benchmark the atomic data in plasma spectral models
\citep{bd2000}.  Work to date on the Capella HETGS spectra has focused
on the identification of strong lines for which HEG and MEG give good
agreement \citep[e.g.,][]{2001ApJ...548..966B, 2001ApJ...549..554A,
2000ApJ...539L..41C}, while we aim to identify weak lines and test a
comprehensive model.

Atomic data for the He-like diagnostic lines themselves are
exceptionally good --- many rates have been benchmarked in controlled
laboratory experiments.  Tokamak data confirm the calculation of the \ion{Ne}{9}
$R$-ratio \citep{coffey94}. The transition probability for the
forbidden line, which determines the density-sensitivity of the \ion{Ne}{9}
$R$-ratio, has been measured to 1\% accuracy on an electron beam ion
trap (EBIT) \citep{wargelin93a}; it has also been confirmed by Bragg
crystal spectrometer measurements on the EBIT \citep{wargelin93b}.
Smith et al. (2001) showed that the largest systematic uncertainties
in the theory for \ion{O}{7} come from the limited number of energy
levels used in calculating the cascades following dielectronic
recombination.

APEC models for \ion{Ne}{9} lines are in good agreement with the
laboratory measurements.  Simplifying assumptions often found in the
literature, such as the lack of temperature-sensitivity in the
$R$-ratio, are no longer necessary since the APEC models are
calculated over dense temperature and density grids.

For weak diagnostic lines such as the \ion{Ne}{9} intercombination
line, two challenging line measurement issues---determining the
continuum level and assessing line blending---are greatly aided by
APEC. The APEC line list includes approximately 40 lines with
reference wavelengths between 13.35 and 13.85\,\AA. Furthermore, the
APEC emissivity table contains $\sim$ 1000 lines in that region, down to 4 orders of
magnitude weaker than the strong resonance line of \ion{Ne}{9}. This
list includes \ion{Ne}{9} dielectronic recombination satellite lines
and Fe and Ni L-shell lines with principal quantum number $n \leq
5$. APEC is sufficiently complete over the HETGS bandpass that it can
be used to select line-free spectral regions for continuum fitting.

\section{Observations and Data Reduction}
\label{obs}

Our approach is to use the six observations of HEG calibration data of
Capella to obtain a very long total exposure time (154.685\,ks) and
benchmark the APEC models in the wavelength region around 13.6\,\AA.
The Capella observations discussed in this paper are summarized in
Table~\ref{obsids}. For each of the three grating instruments, the
data from several observations have been combined to produce high
signal-to-noise spectra. Standard pipeline processing for HETGS,
LETGS, and RGS are described in \cite{2000ApJ...539L..41C},
\cite{2001A&A...365L.324B}, and \cite{2001A&A...365L.329A},
respectively.

\begin{deluxetable*}{crrrccc}
\tabletypesize{\footnotesize}
\tablecolumns{8} 
\tablewidth{0pc} 
\tablecaption{\label{obsids}List of Observations} 
\tablehead{ 
\colhead{Obs\,ID} & \multicolumn{1}{c}{Exp Time} & \multicolumn{2}{c}{Counts\tablenotemark{a}} &
\colhead{Observation Start - End} & \multicolumn{2}{c}{log L$_X$\tablenotemark{b}}\\
\colhead{} & \multicolumn{1}{c}{(ks)} &  &  & \colhead{(YYYY-MM-DDThh:mm:ss)}  & \multicolumn{2}{c}
{[erg s$^{-1}$]}  } 
\startdata
Chandra HETGS & & HEG & MEG & & HEG & MEG \\
00057&28.827&11714&47549&2000-03-03T16:28:53 - 2000-03-04T01:18:03&30.10&30.19\\
01010&29.541&10773&42907&2001-02-11T12:22:49 - 2001-02-11T21:20:17&30.06&30.12\\
01099&14.565&5934&23634&1999-08-28T07:53:17 - 1999-08-28T12:16:37&30.12&30.20\\
01103&40.479&18109&73303&1999-09-24T06:09:21 - 1999-09-24T18:22:56&30.16&30.24\\
01235&14.571&5916&24177&1999-08-28T12:16:37 - 1999-08-28T16:39:57&30.12&30.21\\
01318&26.701&11133&46338&1999-09-25T13:26:39 - 1999-09-25T21:51:17&30.14&30.22\\
\tableline
HETGS Sum&154.685&63579&257908&\nodata&30.12&30.20\\
\tableline
Chandra LETGS & & LEG &&& LEG\tablenotemark{c}&\\
01167&15.36&41215&\nodata&1999-09-09T13:10:06 - 1999-09-09T17:26:08    &30.45&\nodata\\
01244&12.37&32981&\nodata&1999-09-09T17:42:27  -  1999-09-09T21:08:36  &30.45&\nodata\\
01246&15.00&41701&\nodata&1999-09-10T03:06:06  -  1999-09-10T07:16:08  &30.47&\nodata\\
01248&85.36&240047&\nodata&1999-11-09T13:42:24  -  1999-11-10T13:25:05 &30.43&\nodata\\
01420&30.30&83533&\nodata&1999-10-29T22:49:29  -  1999-10-30T07:14:27  &30.44&\nodata\\
62410&11.33&31491&\nodata&1999-09-09T23:43:57  -  1999-09-10T02:52:48  &30.48&\nodata\\
62422&11.68&31202&\nodata&1999-09-12T18:26:42  -  1999-09-12T21:41:20  &30.46&\nodata\\
62423&14.80&39328&\nodata&1999-09-12T23:37:44  -  1999-09-13T03:44:28  &30.45&\nodata\\
62435&22.34&58201&\nodata&1999-09-06T00:35:40  -  1999-09-06T06:48:01  &30.42&\nodata\\
\tableline
LETGS Sum&218.54&599699&\nodata&\nodata&30.41&\nodata\\
\tableline
XMM-Newton&&RGS1&RGS2&&RGS1&RGS2\\
0121920101\tablenotemark{d}&52.92&169423&186267&2000-03-25T11:36:59 - 2000-03-26T02:53:49&30.22&30.30\\
0121920101\tablenotemark{e}&52.92&45115&44524&2000-03-25T11:36:59 - 2000-03-26T02:53:49&30.29&30.15\\
\tablenotetext{a}{Counts are background subtracted values. The two dispersion directions are
co-added.}
\tablenotetext{b}{L$_X$ was determined within the wavelength range 3
to 20 \AA.}
\tablenotetext{c}{Higher order photons are included without correction.}
\tablenotetext{d}{First order.}
\tablenotetext{e}{Second order.}
\enddata
\end{deluxetable*}

\begin{figure}[!ht]
\resizebox{\hsize}{!}{\includegraphics{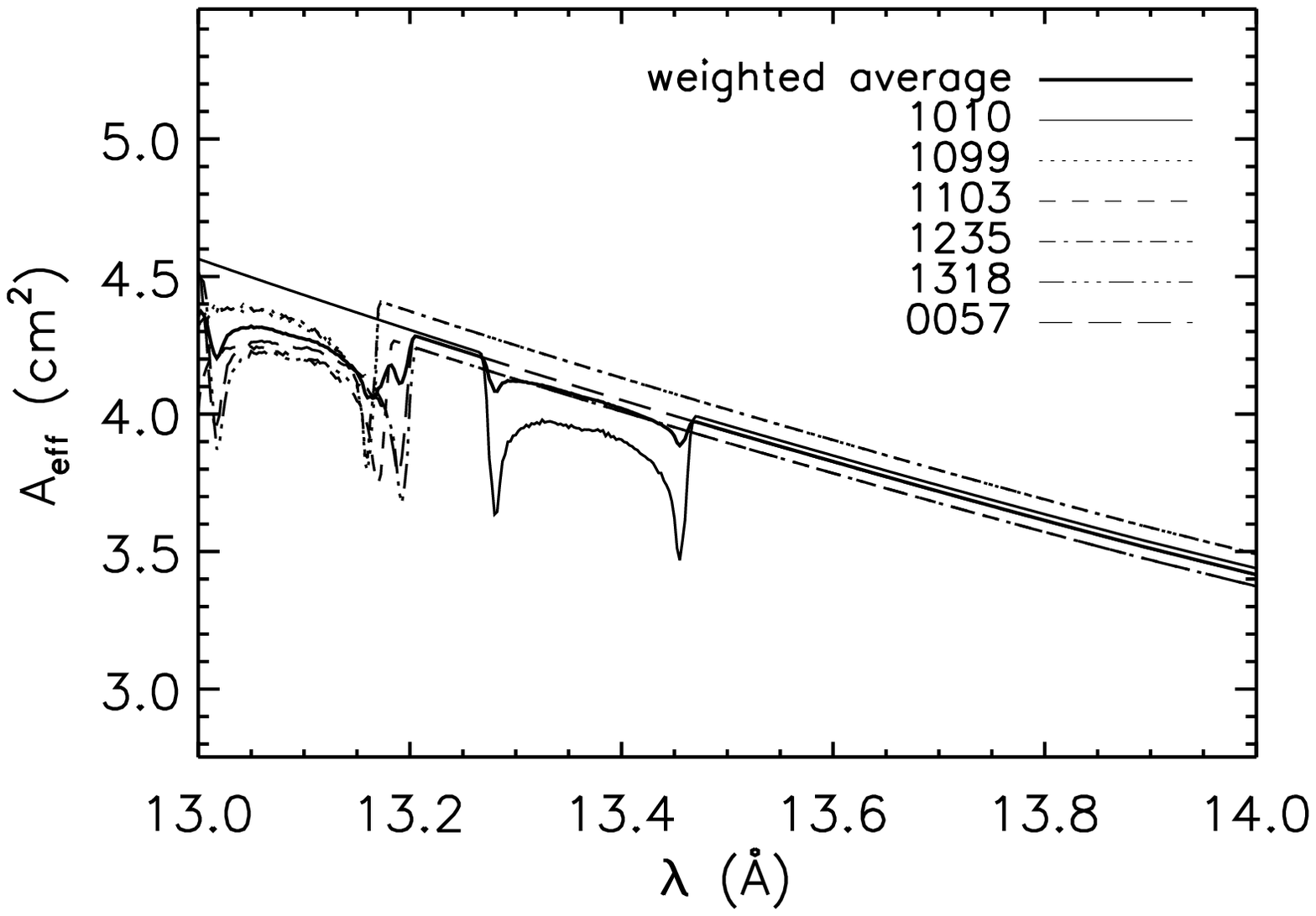}}

\resizebox{\hsize}{!}{\includegraphics{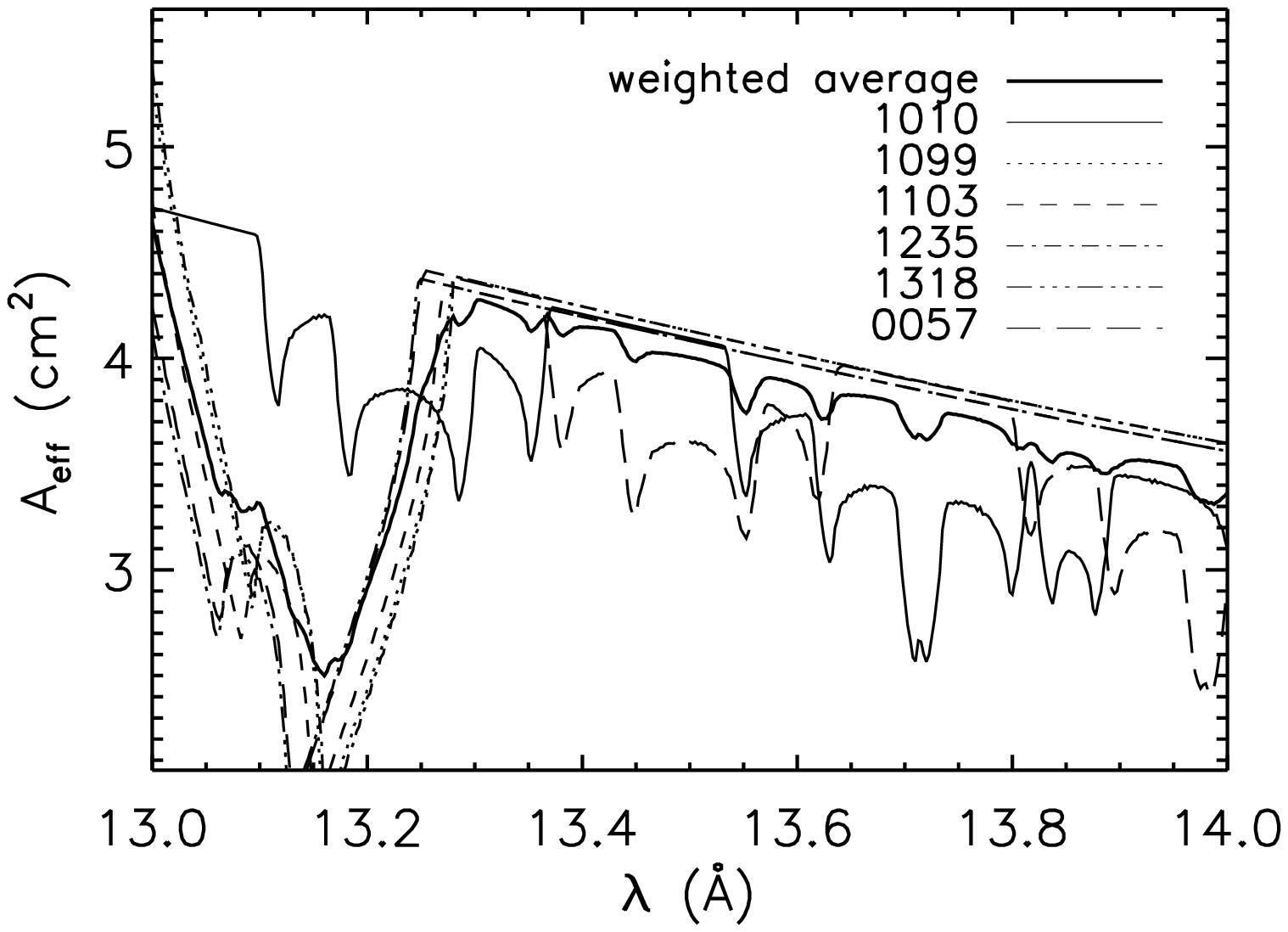}}
\caption{\label{aeff_heg}Effective areas for the different observations
(labeled with observation identification) with the HEG plus first order
(top) and minus first order (bottom). The average is indicated by a thick
line and is obtained by weighting with the exposure times. The large dip on
the short wavelength end of the minus first order is the chip gap, while smaller
dips are caused by the removal of bad pixels. The widths and shapes of these dips
are determined by the aspect dither and the chip geometry.}
\end{figure}

The HETGS/ACIS-S data were obtained from the Chandra archive and
reprocessed with CIAO software Version 2.2\footnote{Available at
http://cxc.harvard.edu/ciao.} and calibration database CALDB Version
2.10. The standard technique for separating the different grating
orders employs a variable-width CCD pulse-height
filter,\footnote{Encoded in the calibration database as the Order
Sorting and Integrated Probability, or OSIP file.} that follows the
pulse-height versus wavelength relation for each spectral order.
Since some observations were made during times of uncertain or
changing detector response (primarily CCD temperature changes that
affected the gain), the standard filter regions did not trace the
actual distributions of events in the pulse-height-wavelength plane
very accurately.  To mitigate this, we used the constant fractional
energy width option available in the CIAO event-resolution program
({\tt tg\_resolve\_events}) with a fractional width of 0.3.  The
constant width region accepts more inter-order background,
particularly at short wavelengths, but this is negligible for our
purposes because of the very low background level.  We have also
bypassed the step of position randomization of events detected within
a given pixel, which is part of the standard image processing of
ACIS-S.  Otherwise, we use the standard CIAO pipeline extraction.  We
use only the first dispersion orders.  The effective areas for
positive and negative first orders have been computed for each dataset
using observation-specific data to account for aspect and bad pixels.
The HEG effective areas for the individual observations in the
wavelength range around 13 \AA\ are shown in Figure~\ref{aeff_heg},
along with the exposure-weighted average effective area to be used for
analysis of the combined spectra.

The LETGS observations and data reduction are described in
\cite{ness01}. They consist of nine datasets added with a total
exposure time of 218.54\,ks. Effective areas are from 2001 February
(see Pease et al.\ 2000 for a preliminary description).  The RGS
observations were taken on 2000 March 25, with an exposure time of
52.92\,ks. Data processing is described in \cite{2003A&A...398.1137A}.  Effective
areas have been calculated for the observations with SAS5.3.

We derive X-ray luminosities given in Table~\ref{obsids} of $L_X = 1.3
\times 10^{30}$, $2.8 \times 10^{30}$, and $1.7 \times
10^{30}$\,ergs\,s$^{-1}$, using the distance of $12.94 \pm 0.15$\,pc
\citep{1997A&A...323L..49P}. Differences are consistent with the
different passbands and responses of the
instruments. Table~\ref{obsids} also shows that the luminosities for
HETGS and LETGS are fairly constant with time (for HETGS over 1.5
years), supporting our comparison of spectra taken with different
instruments at different times.

\section{Data Analysis}

\subsection{Line Flux Measurements}
\label{hegobs}

\begin{deluxetable*}{ccccccccc}
\tabletypesize{\footnotesize}
\tablecolumns{9}
\tablewidth{0pc}
\tablecaption{\label{tab2}Best-fit Line Fluxes for Separate Dispersion Orders of HEG}
\tablehead{
\colhead{$\lambda$}  & \colhead{Amplitude\tablenotemark{a}} & \colhead{Flux\tablenotemark{b}} &
\colhead{A$_{\rm eff}$} & \colhead{} & \colhead{$\lambda$}  & \colhead{Amplitude\tablenotemark{a}} & \colhead{Flux\tablenotemark{b}} & \colhead{A$_{\rm eff}$}\\
\colhead{(\AA)}  & \colhead{(cts)} & \colhead{(ph cm$^{-2}$
ks$^{-1}$)} & \colhead{(cm$^2$)}& \colhead{} & \colhead{(\AA)}  & \colhead{(cts)} &
\colhead{(ph cm$^{-2}$ ks$^{-1}$)} & \colhead{(cm$^2$)} }
\startdata
\multicolumn{4}{l}{+1$^{\rm st}$ order} & & \multicolumn{4}{l}{-1$^{\rm st}$ order}\\
13.449 $\pm$ 0.002& 315.9 $\pm$  18.4 &  0.511 $\pm$  0.030& 4.00 & &13.449 $\pm$ 0.003& 300.6 $\pm$  17.9 &  0.474 $\pm$  0.028& 4.10 \\
13.553 $\pm$ 0.008& 101.7 $\pm$  10.8 &  0.169 $\pm$  0.018& 3.88 & &13.555 $\pm$ 0.006& 89.9  $\pm$  10.1 &  0.148 $\pm$  0.017& 3.92 \\
13.700 $\pm$ 0.004& 190.7 $\pm$  14.4 &  0.331 $\pm$  0.025& 3.72 & &13.700 $\pm$ 0.005& 182.8 $\pm$  14.0 &  0.312 $\pm$  0.024& 3.78 \\
13.358 $\pm$ 0.021& 31.1  $\pm$   6.4 &  0.050 $\pm$  0.010& 4.07 & &13.356 $\pm$ 0.020& 31.1  $\pm$   6.4 &  0.048 $\pm$  0.010& 4.21 \\
13.381 $\pm$ 0.007& 40.6  $\pm$   7.2 &  0.065 $\pm$  0.012& 4.04 & &13.378 $\pm$ 0.007& 39.5  $\pm$   7.0 &  0.061 $\pm$  0.011& 4.17 \\
13.404 $\pm$ 0.007& 52.0  $\pm$   8.0 &  0.084 $\pm$  0.013& 3.99 & &13.404 $\pm$ 0.007& 60.2  $\pm$   8.4 &  0.097 $\pm$  0.013& 4.02 \\
13.427 $\pm$ 0.009& 64.0  $\pm$   8.7 &  0.103 $\pm$  0.014& 4.01 & &13.426 $\pm$ 0.007& 74.8  $\pm$   9.3 &  0.117 $\pm$  0.015& 4.13 \\
13.469 $\pm$ 0.004& 134.8 $\pm$  12.5 &  0.219 $\pm$  0.020& 3.98 & &13.468 $\pm$ 0.004& 145.3 $\pm$  12.9 &  0.236 $\pm$  0.021& 3.98 \\
13.507 $\pm$ 0.002& 193.6 $\pm$  15.0 &  0.318 $\pm$  0.025& 3.93 & &13.507 $\pm$ 0.002& 204.0 $\pm$  15.7 &  0.330 $\pm$  0.025& 4.00 \\
13.524 $\pm$ 0.002& 316.0 $\pm$  18.7 &  0.522 $\pm$  0.031& 3.91 & &13.523 $\pm$ 0.001& 261.7 $\pm$  17.4 &  0.425 $\pm$  0.028& 3.98 \\
13.650 $\pm$ 0.008& 71.2  $\pm$   9.1 &  0.122 $\pm$  0.016& 3.78 & &13.646 $\pm$ 0.007& 70.8  $\pm$   9.2 &  0.119 $\pm$  0.015& 3.85 \\
13.675 $\pm$ 0.009& 55.8  $\pm$   8.3 &  0.096 $\pm$  0.014& 3.75 & &13.673 $\pm$ 0.009& 42.1  $\pm$   7.4 &  0.073 $\pm$  0.013& 3.75 \\
13.722 $\pm$ 0.006& 54.7  $\pm$   8.2 &  0.096 $\pm$  0.014& 3.70 & &13.722 $\pm$ 0.008& 47.7  $\pm$   7.7 &  0.082 $\pm$  0.013& 3.76 \\
13.740 $\pm$ 0.008& 42.2  $\pm$   7.3 &  0.074 $\pm$  0.013& 3.68 & &13.742 $\pm$ 0.008& 50.2  $\pm$   7.8 &  0.089 $\pm$  0.014& 3.66 \\
13.779 $\pm$ 0.004& 98.4  $\pm$  10.9 &  0.175 $\pm$  0.019& 3.64 & &13.778 $\pm$ 0.007& 94.3  $\pm$  10.5 &  0.166 $\pm$  0.019& 3.67 \\
13.797 $\pm$ 0.003& 146.6 $\pm$  12.9 &  0.262 $\pm$  0.023& 3.62 & &13.797 $\pm$ 0.006& 117.1 $\pm$  11.4 &  0.205 $\pm$  0.020& 3.69 \\
13.828 $\pm$ 0.004& 185.8 $\pm$  14.1 &  0.335 $\pm$  0.025& 3.59 & &13.826 $\pm$ 0.003& 172.0 $\pm$  13.8 &  0.313 $\pm$  0.025& 3.55 \\
13.847 $\pm$ 0.007& 50.4  $\pm$   8.0 &  0.091 $\pm$  0.014& 3.57 & &13.843 $\pm$ 0.006& 55.4  $\pm$   8.5 &  0.102 $\pm$  0.016& 3.50\\

\tablenotetext{a}{\small Measured line counts with 1$\sigma$ errors. The line
widths are all 0.005\,\AA\ and the assumed continuum level (including
background) for each dispersion order is  350\,cts\,\AA$^{-1}$.}
\tablenotetext{b}{The total exposure time was 154.7~ks.}
\enddata
\end{deluxetable*}


\begin{figure}[!ht]
\resizebox{\hsize}{!}{\includegraphics{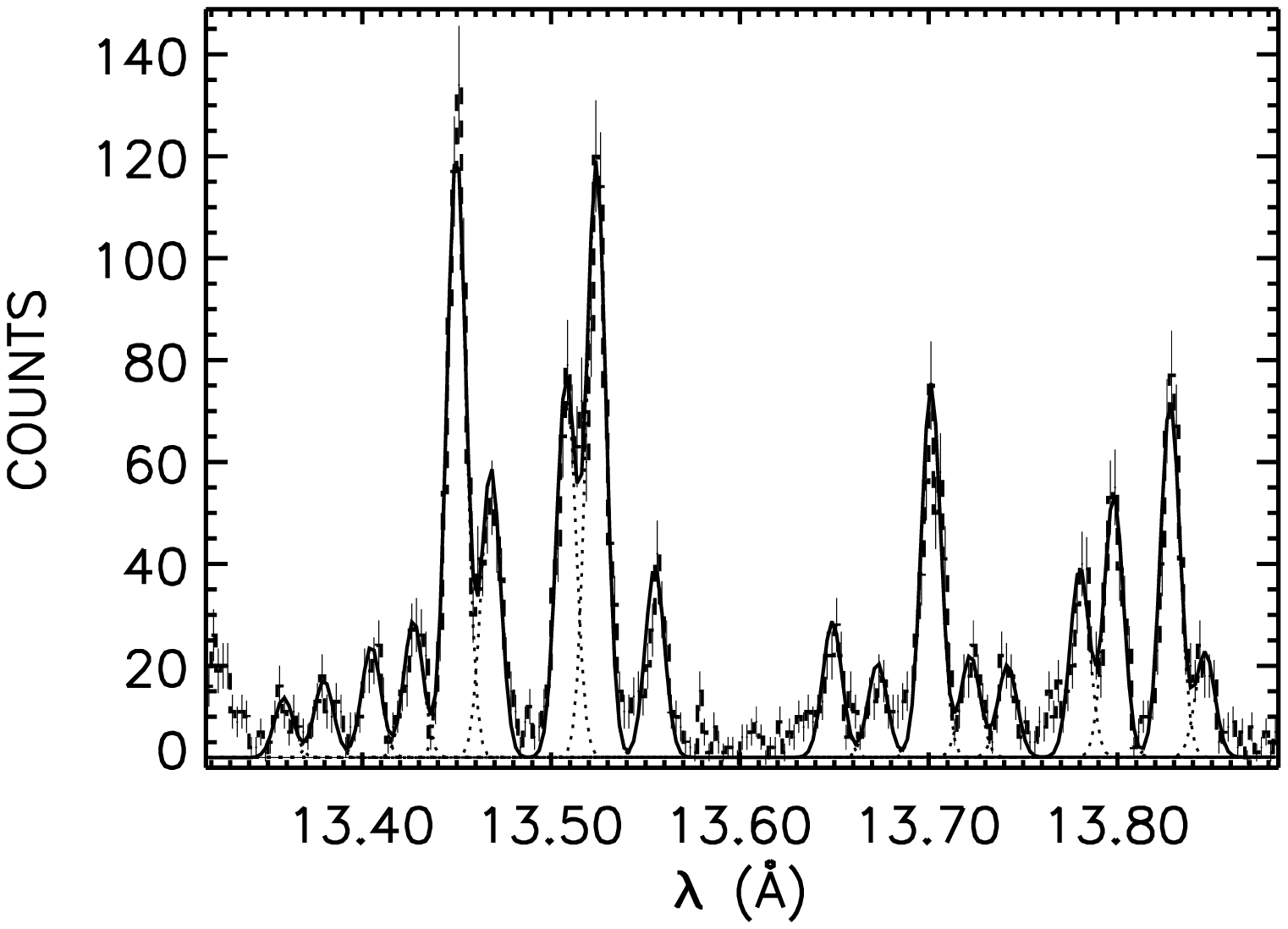}}
\caption{\label{hegfit}
The HEG summed spectrum of Capella using plus and minus first order. In order to construct an empirical model, eighteen emission lines are
used with Gaussian line widths $\sigma=0.005$\,\AA\ (equivalent to
FWHM$= 0.012$\,\AA). The continuum level (including weak lines and
background, for combined plus and minus first orders) is chosen
constant with 700\,cts\,\AA$^{-1}$. The binsize is 2.5\,m\AA\ and the
exposure time is 154.7\,ks.}
\end{figure}

Table~\ref{tab2} lists 18 strong emission line features measured in
the plus and minus first order HEG spectra between 13.35 and
13.85\,\AA, including several features not previously reported for
Capella. Line counts are measured with the CORA line fitting tool
developed by \cite{ness01} and described and refined by
\cite{newi}. The raw counts obtained with CORA represent the fitted
number of expected counts for a given continuum (plus background),
requiring Poissonian statistics to be conserved. Measurement errors
are given as $1\,\sigma$ errors and include statistical errors and
correlated errors in cases of line blends, but include no systematic
errors from, e.g., the continuum placement. Line widths are fixed rather
than fit, since line broadening is expected to be predominantly
instrumental and line profiles are represented by analytical profile
functions (Gaussians for HEG and MEG, Lorentzians for RGS, and 
for the LETGS modified Lorentzians  $F(\lambda)=a/(1+\frac{\lambda-\lambda_0}{\Gamma})^\beta$ with
$\beta=2.5$; \cite{kashyap}). The continuum is chosen
as an initial guess to be a constant with level determined by the HEG
region at 13.6\,\AA, which contains no apparent lines in the data as
well as no significant emission lines in the APEC database.  (The
placement of the continuum is discussed further in \S\ref{cont}.)

The wavelengths and line counts are listed in Table~\ref{tab2}, along
with fluxes determined using the total exposure time and effective
areas given in the last column for each line. We find the spectra from
both dispersion directions to agree reasonably well and use the plus
and minus first order-summed spectrum for further analysis
(Table~\ref{feid}). The summed spectrum is shown in
Figure~\ref{hegfit} along with the empirical model, i.e., continuum
plus 18 lines, with best-fit centroids and fluxes.

\begin{deluxetable*}{lcccr@{ -- }lccc}
\tabletypesize{\footnotesize}
\tablecolumns{5}
\tablewidth{0pc}
\tablecaption{\label{feid}Line Identification for HEG Measurements\tablenotemark{a}}
\tablehead{
\colhead{$\lambda$\tablenotemark{b}$_{obs}$} &
\colhead{$\lambda$\tablenotemark{c}$_{ref}$} &
\colhead{$\lambda$\tablenotemark{c}$_{err}$} &
\colhead{Ion} &
\multicolumn{2}{c}{Transition} &
\colhead{Flux$_{obs}$} &
\colhead{Flux\tablenotemark{d}$_{model}$}\\
\colhead{(\AA)} &
\colhead{(\AA)} &
\colhead{(\AA)} &
\colhead{} &
\colhead{} &
\colhead{} &
\colhead{(ph cm$^{-2}$ ks$^{-1}$)} &
\colhead{(ph cm$^{-2}$ ks$^{-1}$)} &
\colhead{}
}
\startdata
13.354& 13.355 & 0.009   & Fe\,{\sc xviii}& $2p^5~^2P_{3/2}$    & $2s2p^5(^3P)3p~^2P_{3/2}$        & 0.048$\pm$0.007  &   0.0276 \\
13.377& 13.385 & 0.004   & Fe\,{\sc xx}   & $2s2p^4~^4P_{5/2}$  & $2s2p^3(^5S)3d~^4D_{7/2}$        & 0.062$\pm$0.008 &   0.0577 \\
      & 13.390 & \nodata & Fe\,{\sc xx}   & $2p^3~^2P_{3/2}$    & $2p_{1/2}2p_{3/2}3d_{5/2}$       & \nodata            &   0.0046\\
13.401& 13.395 & \nodata & Fe\,{\sc xviii}& $2p^5~^2P_{3/2}$    & $2s2p^5(^3P)3p~^2D_{5/2}$        & 0.091$\pm$0.009 &   0.0459 \\
      & 13.407 & 0.004   & Fe\,{\sc xviii}& $2p^5~^2P_{3/2}$    & $2s2p_{1/2}^22p_{3/2}^33p_{3/2}$ & \nodata            &   0.0238\\
      & 13.409 & \nodata & Fe\,{\sc xx}   & $2s2p^4~^4P_{5/2}$  & $2s2p_{1/2}^22p_{3/2}3d_{5/2}$   & \nodata            &   0.0169\\
      & 13.418 & \nodata & Fe\,{\sc xx}   & $2s2p^4~^4P_{3/2}$  & $2s2p_{1/2}2p_{3/2}^23s$         & \nodata            &   0.0108\\
13.424& 13.423 & 0.004   & Fe\,{\sc xix}  & $2p^4~^3P_{2}$      & $2p^3(^2D)3d~^1F_{3}$            & 0.110$\pm$0.010 &   0.0504 &\\
            & 13.431 & \nodata & Fe\,{\sc xxi}  & $2s2p^3~^3S_{1}$    & $2s2p_{1/2}2p_{3/2}3s$           & \nodata            &   0.0080 \\
            & 13.432 & \nodata & Fe\,{\sc xxi}  & $2p^3~^2P_{1/2}$    & $2p^2(^3P)3d~^2P_{3/2}$          & \nodata            &   0.0042 \\
13.446& 13.447 & 0.004   & Ne\,{\sc ix}   & $1s^2~^1S_{0}$      & $1s2p~^1P_{1}$                   & 0.492$\pm$0.021 &   0.3978 &\\
13.465& 13.462 & 0.003   & Fe\,{\sc xix}  & $2p^4~^3P_{2}$      & $2p^3(^2D)3d~^3S_{1}$            & 0.228$\pm$0.015 &   0.1145 &\\
13.504& 13.497 & 0.005   & Fe\,{\sc xix}  & $2p^4~^3P_{2}$      & $2p_{1/2}2p_{3/2}^23d_{3/2}$     & 0.325$\pm$0.018 &   0.2008 &\\
      & 13.507 & 0.005   & Fe\,{\sc xxi}  & $2s2p^3~^3D_{1}$    & $2s2p_{1/2}^23s$                 & \nodata            &   0.0579 \\
13.521& 13.518 & 0.002   & Fe\,{\sc xix}  & $2p^4~^3P_{2}$      & $2p^3(^2D)3d~^3D_{3}$            & 0.471$\pm$0.021 &   0.4425 &\\
13.551& 13.550 & 0.005   & Ne\,{\sc ix}   & $1s^2~^1S_{0}$      & $1s2p~^3P_{2}$                   & 0.158$\pm$0.012 &   0.0021 &\\
      & 13.551 & 0.005   & Fe\,{\sc xix}  & $2p^4~^3P_{2}$      & $2p_{1/2}2p_{3/2}^23d_{5/2}$     & \nodata            &   0.0290 &\\
      & 13.553 & 0.005   & Ne\,{\sc ix}   & $1s^2~^1S_{0}$      & $1s2p~^3P_{1}$                   & \nodata            &   0.0531 \\
      & 13.554 & \nodata & Fe\,{\sc xix}  & $2p^4~^3P_{2}$      & $2p_{1/2}^22p_{3/2}3d_{5/2}$     & \nodata            &   0.0101 &\\
      & 13.558 & \nodata & Fe\,{\sc xx}   & $2s2p^4~^4P_{1/2}$  & $2s2p_{1/2}^22p_{3/2}3d_{3/2}$   & \nodata            &   0.0112 \\
13.645& 13.645 & 0.004   & Fe\,{\sc xix}  & $2p^4~^3P_{2}$      & $2p^3(^2D)3d~^3F_{3}$            & 0.118$\pm$0.011 &   0.0712 &\\
      & 13.648 & \nodata & Fe\,{\sc xix}  & $2p^4~^3P_{2}$      & $2p^3(^4S)3d~^3D_{3}$            & \nodata            &   0.0175 \\
13.671& 13.674\tablenotemark{e}
               & \nodata & Fe\,{\sc xix}  & $2p^4~^3P_{1}$      & $2p_{1/2}2p_{3/2}^23d_{5/2}$     & 0.085$\pm$0.010 &   0.0133 &\\
      & 13.675 & \nodata & Fe\,{\sc xix}  & $2p^4~^3P_{1}$      & $2p_{1/2}2p_{3/2}^23d_{3/2}$     & \nodata            &   0.0264 \\
      & 13.683 & \nodata & Fe\,{\sc xix}  & $2p^4~^3P_{2}$      & $2p_{1/2}2p_{3/2}^23d_{3/2}$     & \nodata            &   0.0208 \\
13.697& 13.699 & 0.005   & Ne\,{\sc ix}   & $1s^2~^1S_{0}$      & $1s2s~^3S_{1}$                   & 0.322$\pm$0.017 &   0.1845 \\
13.719& 13.732\tablenotemark{f}
               & \nodata & Fe\,{\sc xix}  & $2p^4~^3P_{1}$      & $2p_{1/2}2p_{3/2}^23d_{5/2}$     & 0.089$\pm$0.010 &   0.0573 &\\
13.738& 13.746 & \nodata & Fe\,{\sc xix}  & $2p^4~^1D_{2}$      & $2p^3(^2D)3d~^1F_{3}$            & 0.081$\pm$0.010 &   0.0690 &\\
13.775& 13.767 & 0.005   & Fe\,{\sc xx}   & $2p^3~^4S_{3/2}$    & $2p^2(^3P)3s~^4P_{5/2}$          & 0.170$\pm$0.013 &   0.0494 &\\
      & 13.779 & 0.005   & Ni\,{\sc xix}  & $2p^6~^1S_{0}$      & $2p^5(^2P)3s~^1P_{1}$            & \nodata            &   0.0676 \\
13.794& 13.795 & 0.005   & Fe\,{\sc xix}  & $2p^4~^3P_{2}$      & $2p_{1/2}2p_{3/2}^23d_{5/2}$     & 0.234$\pm$0.015 &   0.1778 \\
      & 13.795 & 0.005   & Fe\,{\sc xix}  & $2p^4~^1D_{2}$      & $2p^3(^2D)3d~^3P_{2}$            & \nodata            &   0.0207 \\
13.824& 13.825 & 0.002   & Fe\,{\sc xvii} & $2p^6~^1S_{0}$      & $2s2p^63p~^1P_{1}$               & 0.325$\pm$0.018 &   0.3058 &\\
      & 13.839 & 0.005   & Fe\,{\sc xix}  & $2p^4~^3P_{2}$      & $2p_{1/2}^22p_{3/2}3d_{5/2}$     & \nodata            &   0.0321 \\
      & 13.839 & 0.005   & Fe\,{\sc xix}  & $2p^4~^1D_{2}$      & $2p_{1/2}^22p_{3/2}3d_{5/2}$     & \nodata            &   0.0082 \\
13.843& 13.843 & 0.006   & Fe\,{\sc xx}   & $2p^3~^4S_{3/2}$    & $2p^2(^3P)3s~^4P_{3/2}$          & 0.094$\pm$0.011 &   0.0233 \\

\tablenotetext{a}{Only the strongest lines modeled are identified. In cases of line blends, weaker lines up to about 10\% of the strong line 
flux are also listed.}
\tablenotetext{b}{Measured HEG wavelengths have been corrected according to the scaling.}
\tablenotetext{c}{Sources for reference wavelengths with errors are given in the text.}
\tablenotetext{d}{Model fluxes are based on the emission measure distribution derived from the lines in 
Table~\ref{em_fe} (shown in Fig.~\ref{emmodel}).}
\tablenotetext{e}{Brown et al. (1999) list a group of 5 lines
(``O21'') with a
central wavelength 13.676 $\pm$ 0.004, although APEC has not assigned
this wavelength to any of the model lines.}
\tablenotetext{f}{APEC includes two relatively strong Ne\,{\sc ix}
satellite lines which may contribute to a centroid shift.}
\enddata
\end{deluxetable*}


\begin{figure}[!ht]
\resizebox{\hsize}{!}{\includegraphics[angle=90]{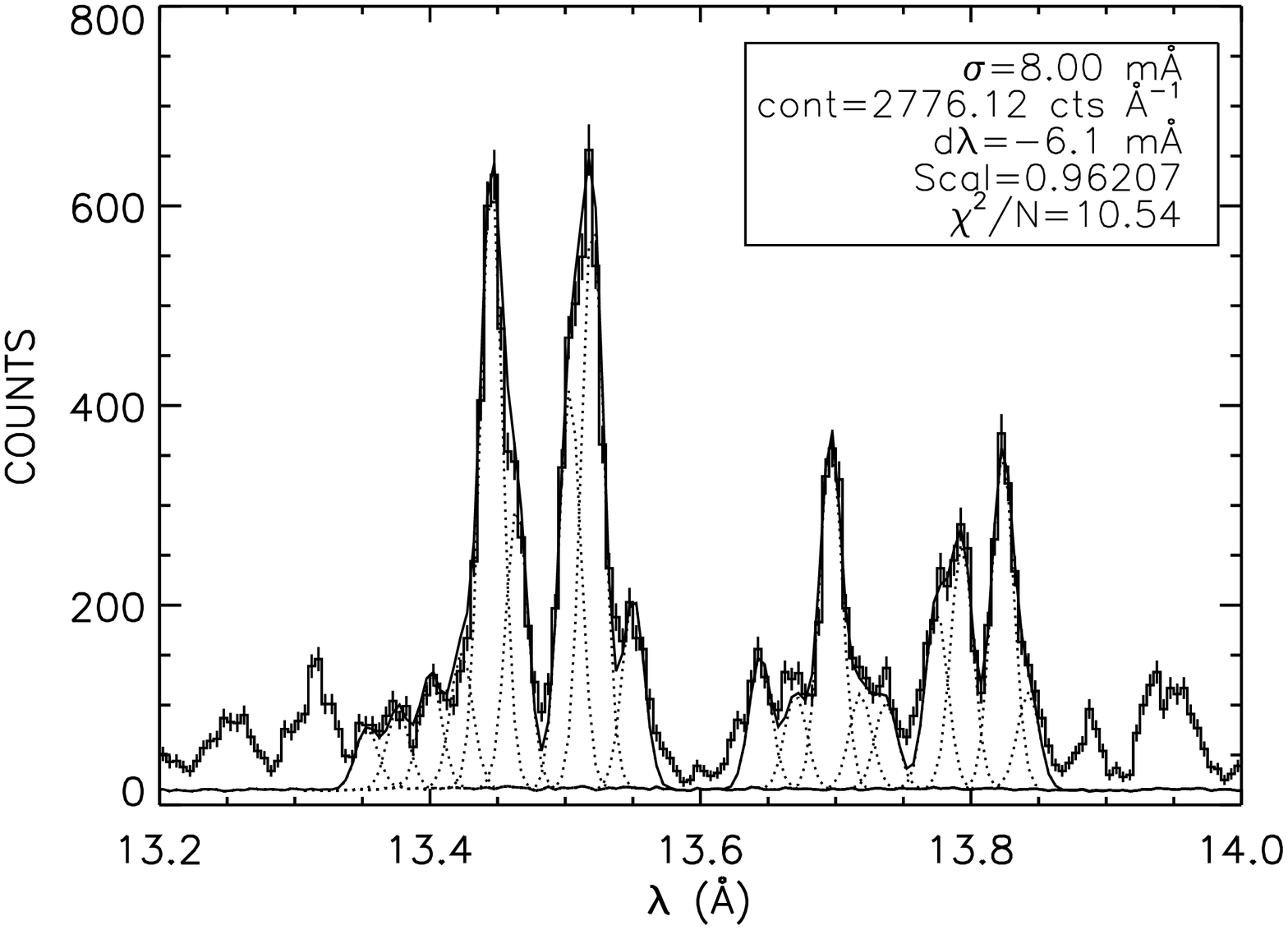}}

\resizebox{\hsize}{!}{\includegraphics[angle=90]{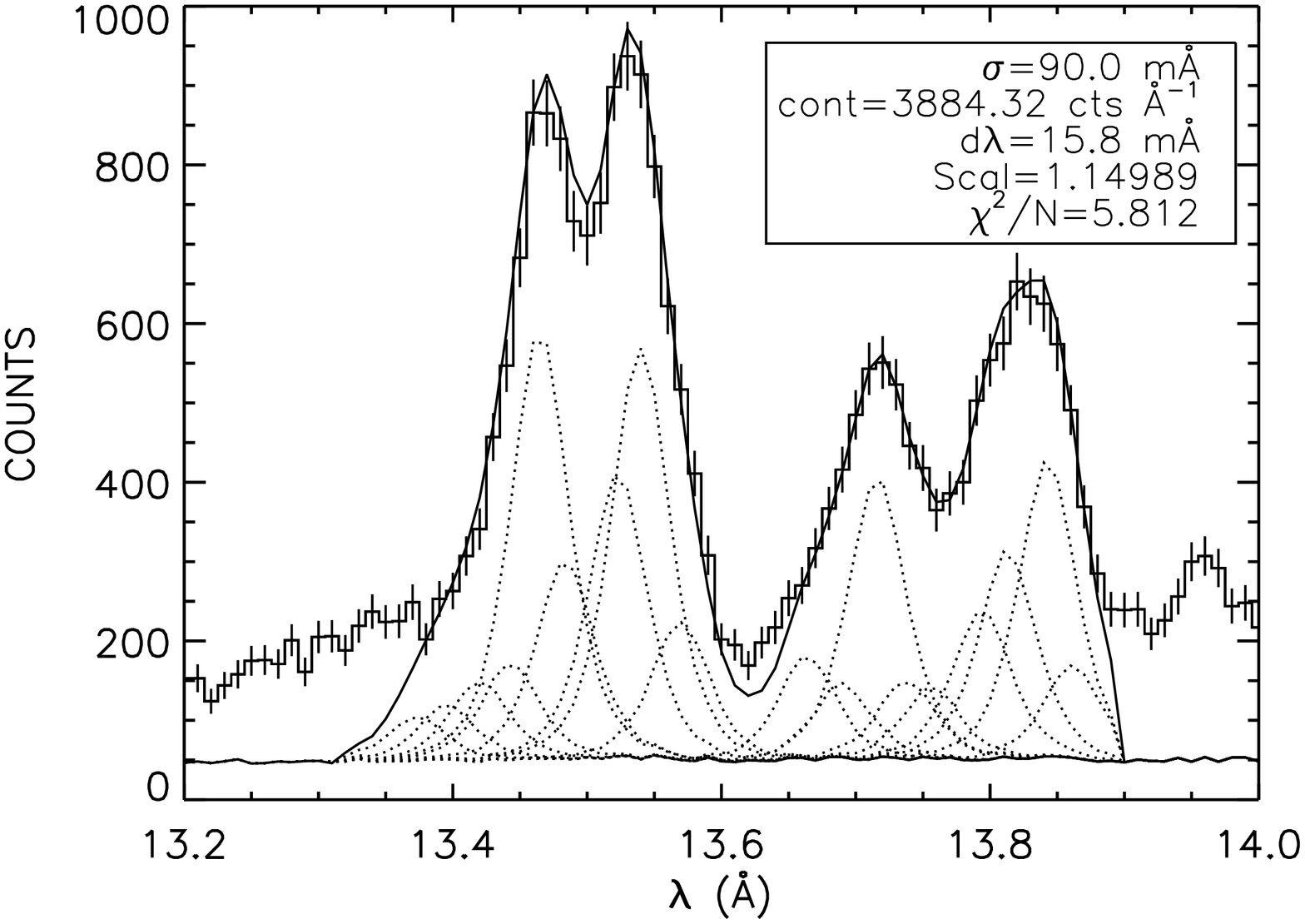}}
\caption{\label{fits}
Best-fit model obtained from the HEG spectrum, scaled and overlaid on
the measured spectra from MEG (top) and LETGS (bottom).  The scaling
parameters d$\lambda$, $Scal$, $\sigma$, and $cont$ (described in the
text) are listed in the upper right box. The goodness of fit
$\chi^2/N$ is also given. For MEG the binsize is 5.0\,m\AA\ and the
exposure time is 154.7\,ks. For LETGS the binsize is 10.0\,m\AA\ and
the exposure time is 218.5\,ks.
}
\end{figure}

\begin{figure}[!ht]
\resizebox{\hsize}{!}{\includegraphics[angle=90]{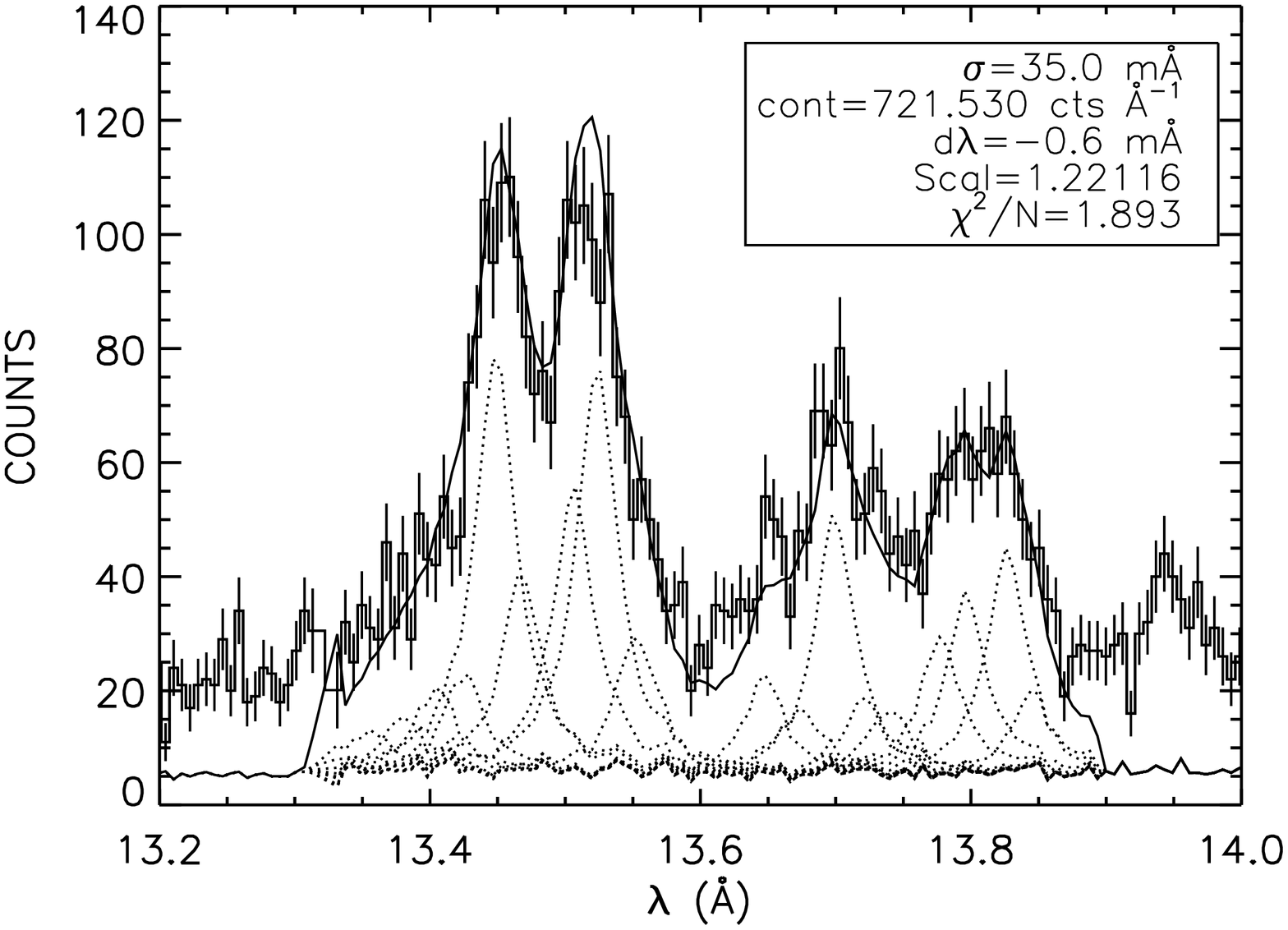}}

\resizebox{\hsize}{!}{\includegraphics[angle=90]{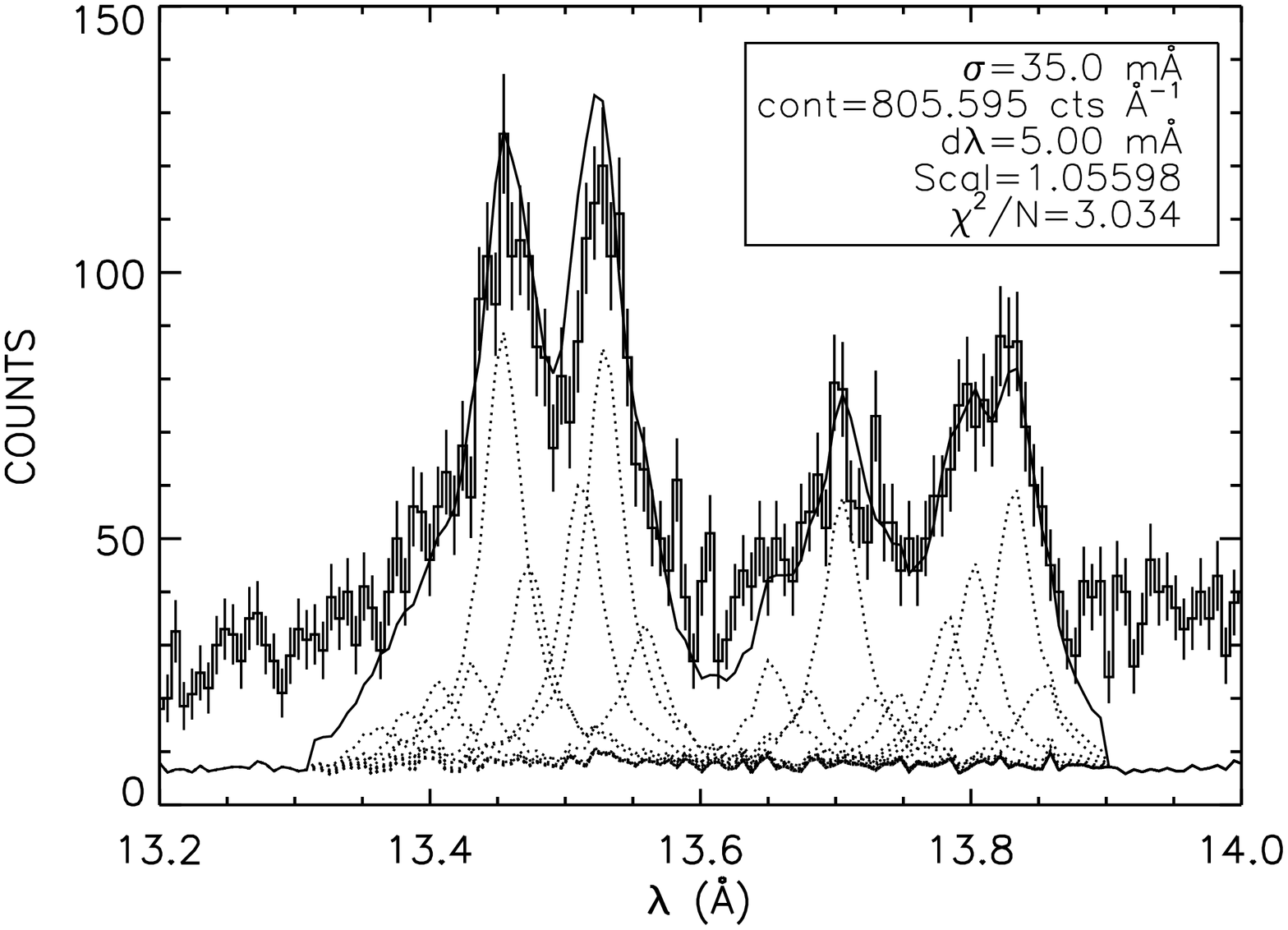}}
\caption{\label{rgssp}Same as Figure~\ref{fits} for the RGS1 (top) and
RGS2 (bottom). For RGS1 the binsize is 5.442\,m\AA\ and for RGS2 the
binsize is 6.048\,m\AA. The exposure time is 52.92\,ks.}
\end{figure}

To predict count spectra from the other instruments, we convolved the
HEG empirical model with each instrumental response.  We allowed
global wavelength shifts (d$\lambda$) to account for different
absolute dispersion calibrations, and a normalization correction
($Scal$) for differences in effective area calibration.  A $\chi^2$
minimization adjusts these two parameters to transform the model to
each instrument.  Figure~\ref{fits} shows the MEG and LETGS model and
data, while Figure~\ref{rgssp} shows RGS1 and RGS2.  The instrumental
line spread function width ($\sigma$) was not a free parameter, since
it is well determined by calibration.  The continuum ($cont$) was also
not a free parameter but was set at the flux level determined from the
HEG spectrum. We find excellent agreement between the observed counts
and the transformed HEG model, within the systematic errors expected
from calibration.

\subsection{Emission Measure Distribution from the HEG Spectrum}
\label{emd}

\begin{figure}[!ht]
\resizebox{\hsize}{!}{\includegraphics{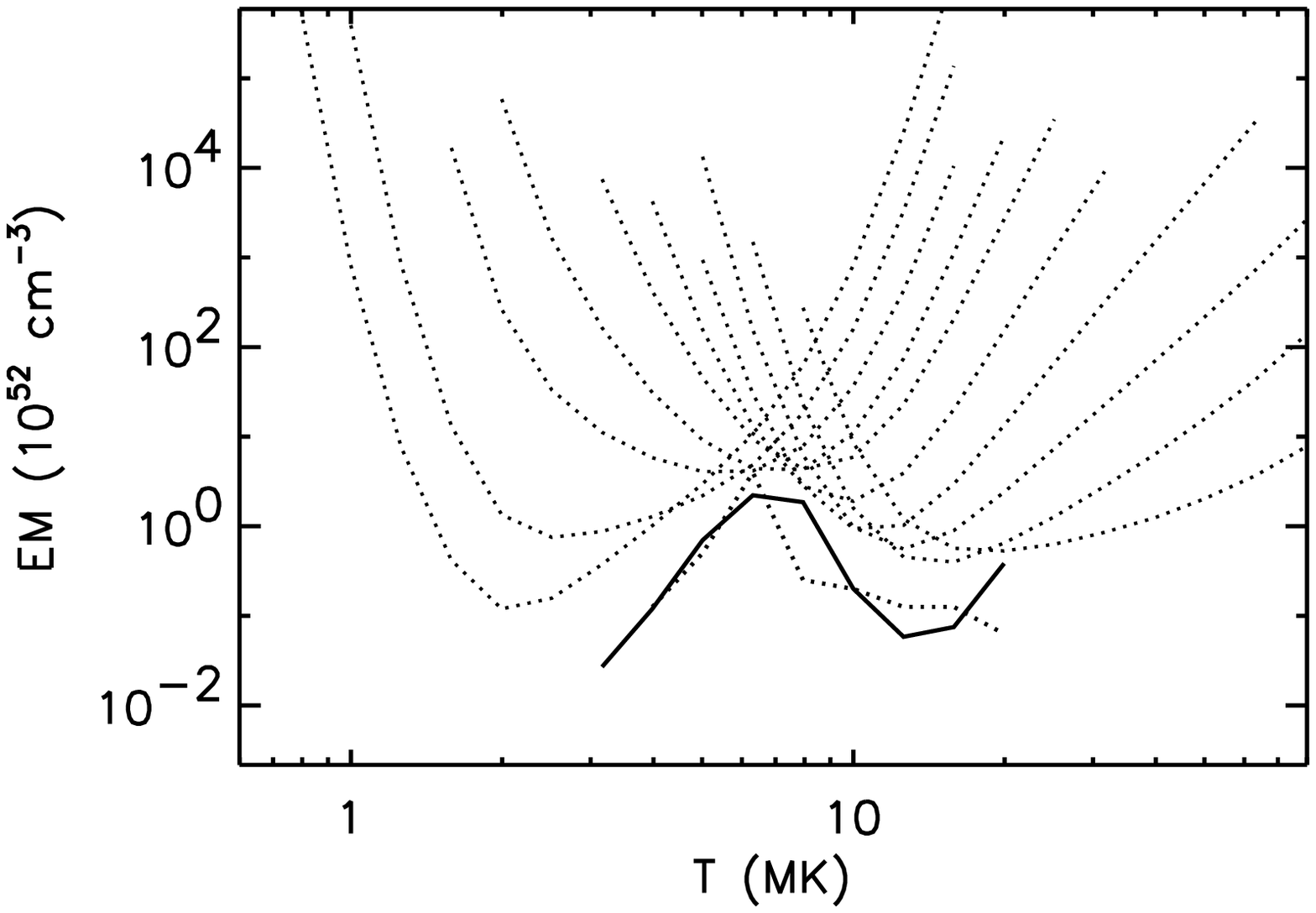}}
\caption[]{\label{emmodel}Best-fit emission measure distribution derived
from iron lines in ionization stages \ion{Fe}{15} to \ion{Fe}{24}
(solid). The result from \cite{2000ApJ...530..387B} is also shown
(thick dotted). Individual line emission measure curves (thin dotted) are shown for the
lines in Table~\ref{em_fe}.}
\end{figure}

In order to compare our measurements with models we construct a rough
emission measure distribution using selected iron lines from
ionization stages \ion{Fe}{15} to \ion{Fe}{24} (see
Table~\ref{em_fe}).  We use the strongest line from each ion (except
for \ion{Fe}{17}) and use other lines to check the model for
consistency.  The EUVE line fluxes are given by
\cite{2001ApJ...549..554A} and are corrected for an interstellar
column density $N_H = 1.8\,\times\,10^{18}$\,cm$^{-2}$
\citep{1993ApJ...402..694L}. Assuming collisional ionization
equilibrium, we construct an emission measure distribution at low
density ($n_e = 1.0$ cm$^{-3}$) using the
APEC line emissivities $\epsilon(T)$.  Each emission line specifies an
emission measure curve $4\pi d^2$ Flux$_{obs}/\epsilon(T)$, which
represents the emission measure as if each temperature contributed the
total emission in the line. It is an upper limit to the final,
self-consistent emission measure distribution determined by using the
lower envelope of the ensemble of curves as the initial estimate in an
iterative optimization to a solution which predicts all measured line
fluxes.  Figure~\ref{emmodel} shows the individual emission measure
curves with the best-fit model. The figure also compares an earlier
emission measure distribution obtained from simultaneous EUVE and ASCA
observations \citep{2000ApJ...530..387B}.

\begin{deluxetable*}{lllr@{ -- }lccc}
\tabletypesize{\small}
\tablewidth{0pc}
\tablecaption{\label{em_fe}HEG Iron Line Measurements Useful for Emission Measure Distribution}
\tablehead{
\colhead{$\lambda_{obs}$} & 
\colhead{$\lambda_{ref}$} & 
\colhead{Ion} & 
\multicolumn{2}{c}{Transition} & 
\colhead{Counts} & 
\colhead{Flux$_{obs}$} & 
\colhead{A$_{\rm eff}$} \\
\colhead{(\AA)} & 
\colhead{(\AA)} & 
\colhead{} & 
\colhead{} & 
\colhead{} & 
\colhead{} & 
\colhead{(ph cm$^{-2}$ ks$^{-1}$)} & 
\colhead{(cm$^2$)} 
}
\startdata
284.15\tablenotemark{a} &  284.16  & Fe\,{\sc xv}    & $3s^2~^1S_{0}$     & $3s3p~^1P_{2}$                   & \nodata       & 10.5\tablenotemark{b}   & \nodata \\
335.41\tablenotemark{a} &  335.41  & Fe\,{\sc xvi}   & $3s~^2S_{1/2}$     & $3p~^2P_{3/2}$                   & \nodata       & 33.9\tablenotemark{b}   & \nodata \\
15.262\tablenotemark{a} &  15.261  & Fe\,{\sc xvii}  & $2p^6~^1S_{0}$     & $2p^5(^2P)3d~^3D_{1}$            & 841.008 $\pm$ 29.68 & 1.128 $\pm$ 0.040    & 4.819 \\
15.015     &  15.014  & Fe\,{\sc xvii}  & $2p^6~^1S_{0}$     & $2p^5(^2P)3d~^1P_{1}$            & 2719.34 $\pm$ 52.71 & 3.354 $\pm$ 0.065    & 5.241 \\
17.098     &  17.096  & Fe\,{\sc xvii}  & $2p^6~^1S_{0}$     & $2p^5(^2P)3s~^3P_{2}$            & 523.964 $\pm$ 22.95 & 2.459 $\pm$  0.108   & 1.377 \\
14.207\tablenotemark{a} &  14.208\tablenotemark{c}  & Fe\,{\sc xviii} & $2p^5~^2P_{3/2}$   & $2p^4(^1D)3d~^2D_{5/2}$      & 1347.92 $\pm$ 37.92 & 1.331 $\pm$  0.037   & 6.547 \\
\nodata    &  14.208\tablenotemark{c}  & Fe\,{\sc xviii} & $2p^5~^2P_{3/2}$   & $2p_{1/2}2p_{3/2}^33d_{5/2}$ &                     &                      &       \\
16.075     &  16.071  & Fe\,{\sc xviii} & $2p^5~^2P_{3/2}$   & $2p^4(^3P)3s~^4P_{5/2}$          & 286.749 $\pm$  17.28 & 0.868 $\pm$  0.052  & 2.135 \\
14.670\tablenotemark{a} &  14.664  & Fe\,{\sc xix}   & $2p^4~^3P_{2}$     & $2p^3(^2D)3s~^3D_{3}$            & 164.356 $\pm$  14.01 & 0.190 $\pm$  0.016  & 5.586 \\
16.110     &  16.110  & Fe\,{\sc xix}   & $2s2p^5~^3P_{2}$   & $2p_{1/2}2p_{3/2}^23p_{1/2}$     & 42.796 $\pm$   7.22  & 0.136 $\pm$  0.023  & 2.037 \\
12.847\tablenotemark{a} &  12.846  & Fe\,{\sc xx}    & $2p^3~^4S_{3/2}$   & $2p_{1/2}2p_{3/2}3d_{3/2}$       & 372.794 $\pm$  20.84 & 0.222 $\pm$  0.012  & 10.875 \\
12.828     &  12.824  & Fe\,{\sc xx}    & $2p^3~^4S_{3/2}$   & $2p_{1/2}2p_{3/2}3d_{3/2}$       & 248.643 $\pm$  18.42 & 0.148 $\pm$  0.011  & 10.882 \\
\nodata    &  12.827  & Fe\,{\sc xx}    & $2p^3~^4S_{3/2}$   & $2p_{1/2}2p_{3/2}3d_{5/2}$       &                      &                     &        \\
\nodata    &  12.822  & Fe\,{\sc xxi}   & $2s2p^3~^3D_{1}$   & $2s2p_{1/2}2p_{3/2}3d_{5/2}$     &                      &                     &        \\
12.286\tablenotemark{a} &  12.284  & Fe\,{\sc xxi}   & $2p^2~^3P_{0}$     & $2p3d~^3D_{1}$                   & 482.550 $\pm$  23.27 & 0.233 $\pm$  0.011  & 13.402 \\
11.771\tablenotemark{a} &  11.770  & Fe\,{\sc xxii}  & $2s^22p~^2P_{1/2}$ & $2s^23d~^2D_{3/2}$               & 177.567 $\pm$  14.80 & 0.077 $\pm$  0.006  & 14.901 \\
12.750     &  12.754  & Fe\,{\sc xxii}  & $2s2p^2~^2D_{3/2}$ & $2s2p_{1/2}3s$                   & 90.500 $\pm$   10.40 & 0.050 $\pm$  0.006  & 11.610 \\
11.741\tablenotemark{a} &  11.736  & Fe\,{\sc xxiii} & $2s2p~^1P_{1}$     & $2s3d~^1D_{2}$                   & 107.723 $\pm$ 12.10  & 0.047 $\pm$  0.005  & 14.666 \\
11.176\tablenotemark{a} &  11.176  & Fe\,{\sc xxiv}  & $2p~^2P_{3/2}$     & $3d~^2D_{5/2}$                   & 90.073 $\pm$   11.32 & 0.032 $\pm$  0.004  & 18.211 \\

\tablenotetext{a}{This line is used to construct the emission measure distribution.}
\tablenotetext{b}{These measurements are from EUVE (Ayres et al. 2001).}\\
\tablenotetext{c}{Sum of both lines in the blend are used in the model.}
\enddata
\end{deluxetable*}


\subsection{Modeling the Continuum}
\label{cont}

Up to this stage of the analysis, the continuum has been taken as a
constant value empirically determined from the spectrum.  Since the
value of the continuum under the weak lines is crucial to the proper
derivation of the $R$-ratio \citep[see][]{brickhouse02}, we discuss
our continuum modeling method in some detail. In this section we
derive a formal result which turns out to be close to our initial
estimate; in general, iteration to subtract the new continuum flux
from lines might be necessary.

In principle, background, weak lines and continuum can be treated
separately in the analysis.  The background, which includes non-source
events from the detector, cosmic rays, as well as source photons
redistributed by mirror scatter, detector aliasing (CCD readout
streaking), and grating scatter, is more than a factor of 20 lower
than the source-model continuum we derive.  The HETGS spectra are not
background-subtracted and so the background is implicitly included in
the continuum model.  However, the HETGS background rejection is very
high: background near 13 \AA\ is estimated\footnote{See the
``Proposers' Observatory Guide,'' Section 8.3,
http://cxc.harvard.edu/proposer/POG.} to be less than 0.017
cts~ks$^{-1}$ \AA$^{-1}$ arcsec$^{-1}$, and the extraction width is 5
arcsec.  Hence, we only expect about 13 cts~\AA$^{-1}$ per order
in the summed HEG spectrum.  Weak lines not directly measured (i.e.,
not listed in Table~\ref{feid}) are treated in \S\ref{weakline} as a
source of systematic uncertainty to the diagnostic line ratios.

The emission measure distribution constructed in \S\ref{emd} was used
to predict both continuum and line fluxes, under the assumption of
solar abundances. The tabulated APEC linelist includes only lines with
emissivities greater than 1.0 $\times$ 10$^{-20}$ ph cm$^3$ s$^{-1}$;
weaker lines are included as a ``pseudo-continuum'' and are added to
the continuum emission spectrum from bremsstrahlung, radiative
recombination, and two-photon emission. The resulting model continuum
is flat between 13 and 14 \AA\ and is about 30\% lower than the
continuum level used in \S\ref{hegobs} (an initial guess chosen to
match at 13.6 \AA), a difference of only about 3
counts per emission line.

This model continuum spectrum was then fit to a set of line-free
spectral regions, defined as spectral bins for which the model line
flux is less than 20\% of the model continuum flux. Both HEG and MEG
were fit jointly using the {\it Sherpa} package in CIAO with the Cash
statistic appropriate for low-count bins \citep{freeman01}. With the
continuum shape fixed by the emission measure distribution, only a
single parameter for the continuum level, or ``normalization,'' was
allowed to vary. The resulting fit continuum level was 6\% higher than
the model continuum level calculated assuming solar abundances. Models with
stricter criteria for determining line-free regions give similar
results. The
continuum level we adopt between 13 and 14\,\AA\ thus seems robust.

Since the emission measure distribution was determined only from iron
lines, its absolute value should be lowered by 6\% to give agreement with the
fit continuum level, requiring a 6\% higher iron-to-hydrogen
abundance ratio. One can draw the preliminary conclusion from this
analysis that the iron-to-hydrogen abundance ratio is close to solar;
determining accurate abundance ratios requires additional checks from
other spectral features and systematic error
analysis which is beyond the scope of this paper. 
						 
\section{Results and Interpretation}

\subsection{Identification of the Emission Lines}
\label{lineid}

\begin{figure*}[ht]
\resizebox{\hsize}{!}{\includegraphics[angle=90]{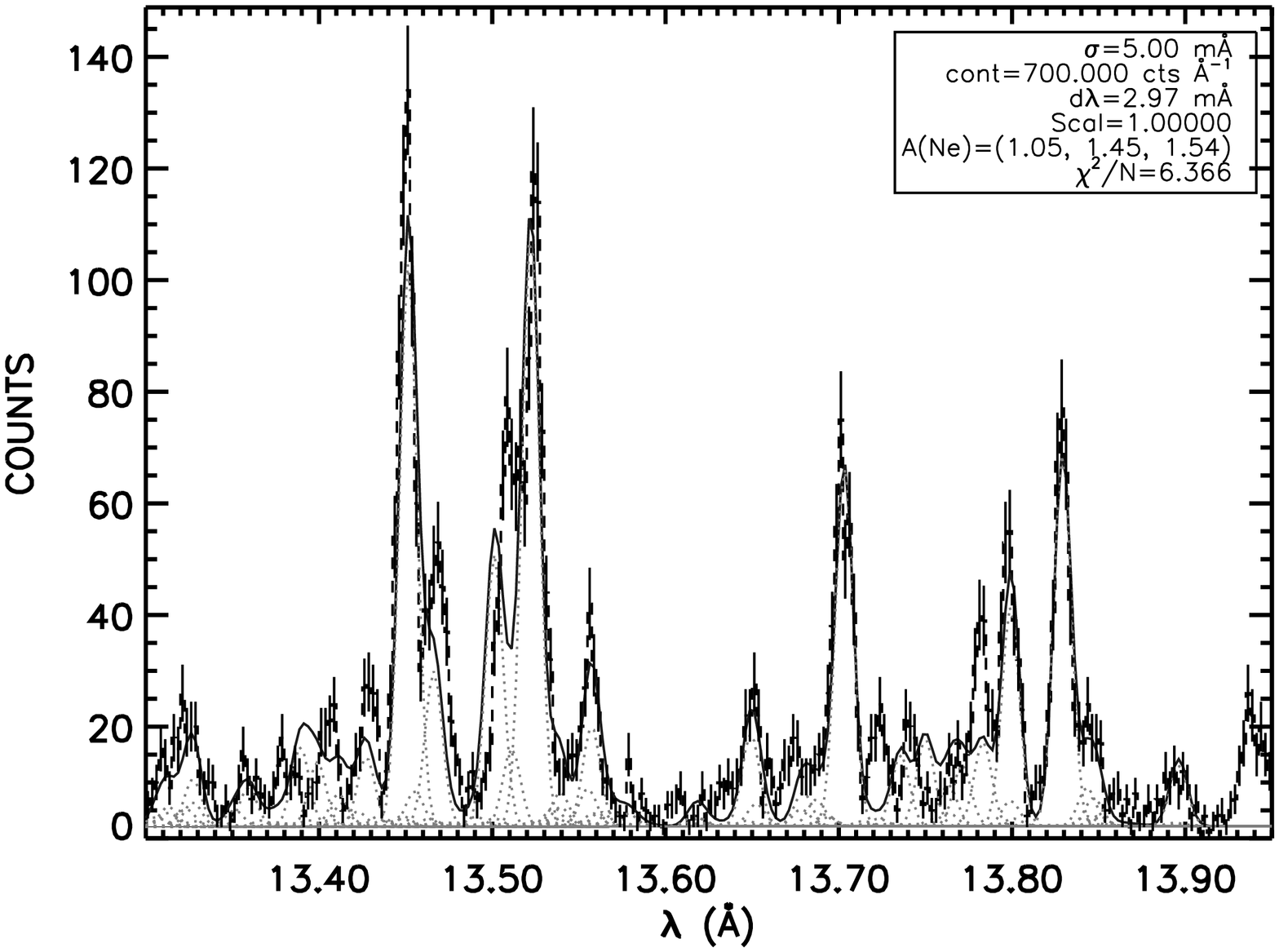}}
\caption[]{\label{modelpred}Measured spectrum of Capella obtained from HEG
  (histogram with error-bars), APEC individual predicted emission
  lines (dotted lines) and their sum (heavy solid line). Predicted
  line fluxes were estimated using the emission measure distribution
  described in \S\ref{emd}. Binsize is 2.5\,m\AA\ and exposure
  time is 154.7 ks.}
\end{figure*}

Using our derived emission measure distribution, we predict the
spectrum of Capella under the assumptions of solar abundances and low
density. We convolve this model spectrum with the
instrumental response, allowing only for the small, systematic, global
shift in wavelengths, and compare it to the observed HEG spectrum. The
model fluxes are given in Table~\ref{feid} along with the measured
fluxes, and the measured spectrum with this model is shown in
Figure~\ref{modelpred}. While the model is calculated at $n_e = 1.0$
cm$^{-3}$, we have used APEC V1.3 density-dependent calculations to
check the sensitivity of our model iron spectrum. For $n_e < 10^8$
cm$^{-3}$, no effects larger than a few percent are found. A few weak
lines in the model show significant density-sensitivity above $10^8$
cm$^{-3}$, but no observable lines show more than 20\% change up to $n_e =
10^{13}$ cm$^{-3}$.

Except for one nickel line, only iron and neon lines are present in
our model of this spectral region.  The most striking discrepancies
are seen for the \ion{Fe}{19} lines at 13.462\,\AA\ (flux
underpredicted by $\sim$ a factor of two), at 13.497\,\AA\ (flux of
blend underestimated by $\sim$ one third, wavelength discrepancy
consistent with errors on laboratory wavelengths), and around 13.72 to
13.75\,\AA\ (wavelength discrepancies well within theoretical errors, no
laboratory wavelengths available). Minor disagreement is seen around
13.4\,\AA. The nickel line at 13.779\,\AA\ is
underpredicted by a factor of two, probably an abundance effect. Since
the goal is to assess blending of the neon diagnostic lines, the
overall level of agreement is a good indicator of our knowledge of the
blending spectrum (especially for \ion{Fe}{19}). We note that: (1) all
strong predicted lines are accounted for in the observed spectrum; (2)
the worst flux disagreement is a factor of two; (3) laboratory iron
wavelengths agree with the observations within the errors, while
theoretical iron wavelengths appear to agree to within one resolution
element of the HEG spectrometer.

Since we have derived the emission
measure distribution from iron lines, we can allow the three neon
diagnostic line fluxes to vary with independent scaling factors,
$A([r,i,f])$, in order to obtain best-fit neon abundances and
densities.  These parameters are iterated with a $\chi^2$
minimization. After applying the scaling factor $A(r)=1.03$ for the
\ion{Ne}{9} resonance line, we find Flux$_{model}$ = 0.41 ph cm$^{-2}$
ks$^{-1}$.
  
\begin{figure}[ht]
\resizebox{\hsize}{!}{\includegraphics[angle=90]{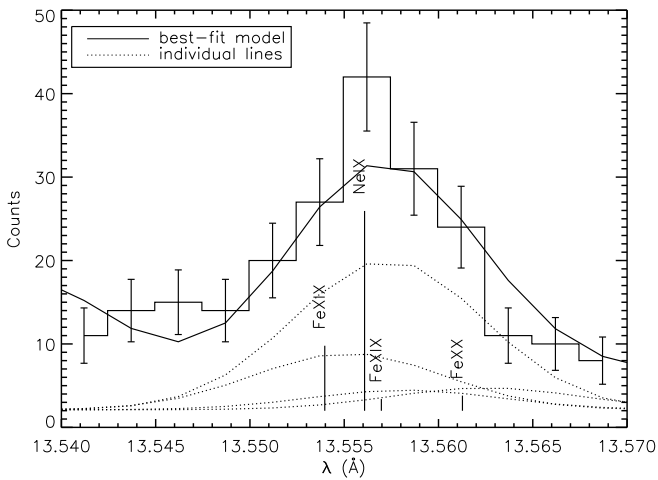}}
\caption[]{\label{iline}Same as Figure~\ref{modelpred} in more detail,
  showing the \ion{Ne}{9} intercombination line with contamination.}
\end{figure}

The intercombination line (with an adjusted flux 0.075 ph cm$^{-2}$
ks$^{-1}$) is significantly weaker in the model than in the observed
spectrum.  This is of course explained by the contaminating lines of
\ion{Fe}{19} and \ion{Fe}{20} that contribute to the observed feature.
Table~\ref{feid} and Figure~\ref{iline} show that the iron lines can
contribute almost half of the total flux measured at 13.55\,\AA, such
that care must be taken with deblending this line even for HEG
spectra. The forbidden line has an adjusted flux 0.28 ph cm$^{-2}$
ks$^{-1}$.

\subsection{Treatment of Weak Line Contamination}
\label{weakline}

Lines which are too weak to be identified and measured in the spectrum
(i.e., lines not found in Table~\ref{feid}) are treated as sources of
systematic error to the diagnostic line fluxes. The total weak line
flux in the region between 13.35 and 13.85\,\AA\ is predicted to be
1.9 times the model continuum flux; however, unlike the flat
continuum, this line emission is not randomly distributed.
Furthermore, most of the stronger of the unidentified lines have
reference wavelengths with small estimated errors. Thus it is
reasonable to compute the fluxes of weak lines within the diagnostic
line profile to estimate the degree of contamination. The models
indicate that the resonance line contamination is $\sim 5$\%, while
the forbidden and intercombination lines each have $\sim 3$\%
contamination. (This is in addition to the contamination from the 50\%
blending of the intercombination line, already discussed.)

\subsection{Density and Temperature Diagnostics with \ion{Ne}{9} for Capella}
\label{dens}

\begin{figure}[!ht]
\resizebox{\hsize}{!}{\includegraphics{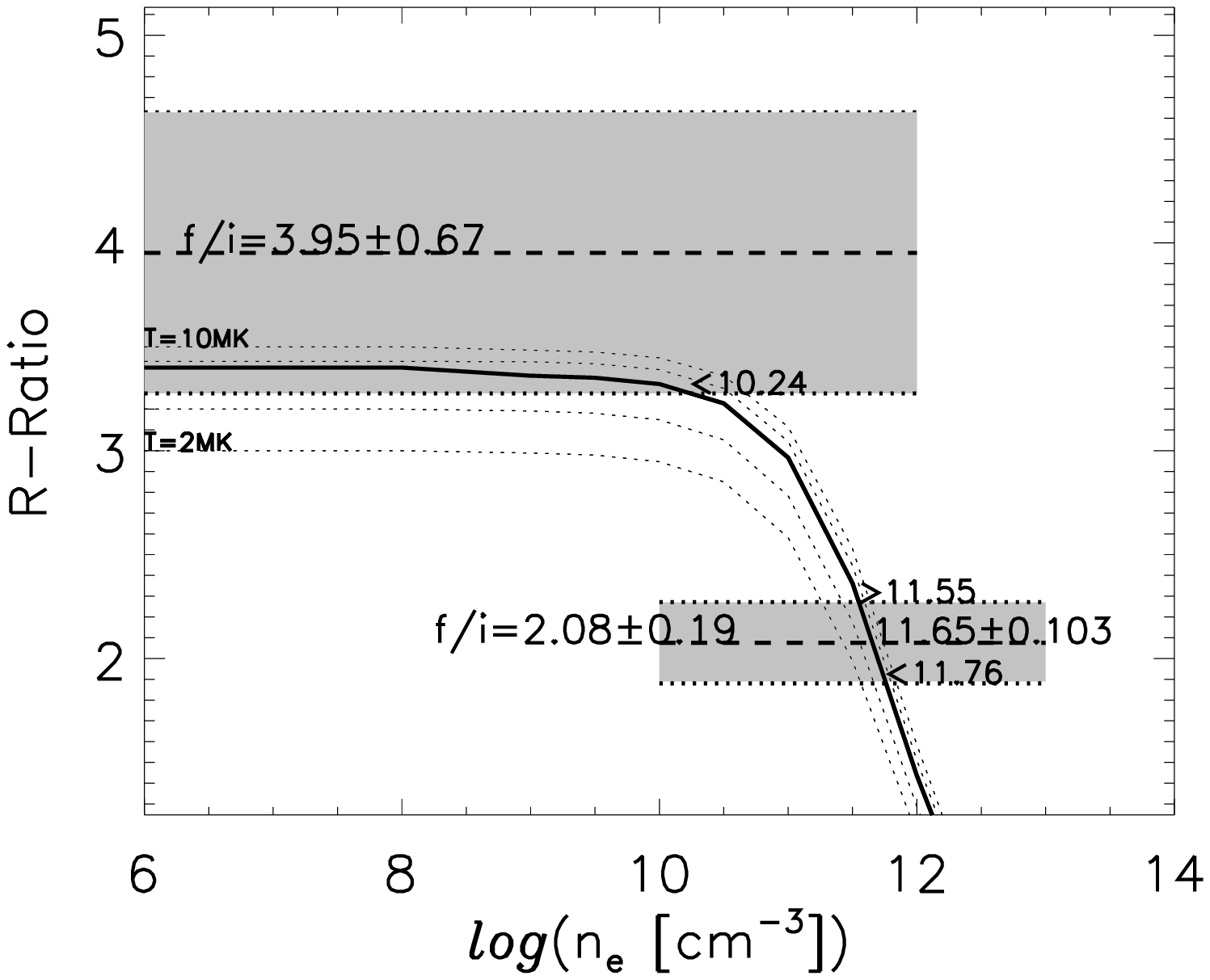}}
\caption{\label{dens_plot}
Determination of the plasma density from the
$R$-ratio using $f$ and $i$ from the direct measurement ($f/i=2.1\,\pm\,0.2$)
and from the model accounting for line blends ($f/i=3.9\,\pm\,0.7$).
Shaded areas represent $1\sigma$ errors on the measured ratios. The upper
limit for the density is $\log n_e < 10.2$. APEC models for the
density-sensitive curves are shown for $T=$ 2.0, 4.0, 6.3, 8.0, and
10 MK, with the curve denoting the 6.3 MK model corresponding to the
peak of the emission measure distribution. The $R$-ratio increases with
temperature.}
\end{figure}

The \ion{Ne}{9} line fluxes constrain electron densities and
temperatures.  When blends are neglected, the Ne-Fe blended
measurement (Table~\ref{feid}) yields an $R$-ratio of $2.1 \pm 0.3$,
which indicates a density $\log (n_e)=11.6\ \pm\ 0.1$.  Instead,
accounting for the blends in the intercombination line based on the
emission measure model (see Fig.~\ref{iline}) we derive the
low-density limit: $R=3.95 \pm 0.7$ resulting in $\log(n_e)<10.2$.
Figure~\ref{dens_plot} shows that the low-density limit is found at
all relevant temperatures.

\begin{figure}[!ht]
\resizebox{\hsize}{!}{\includegraphics{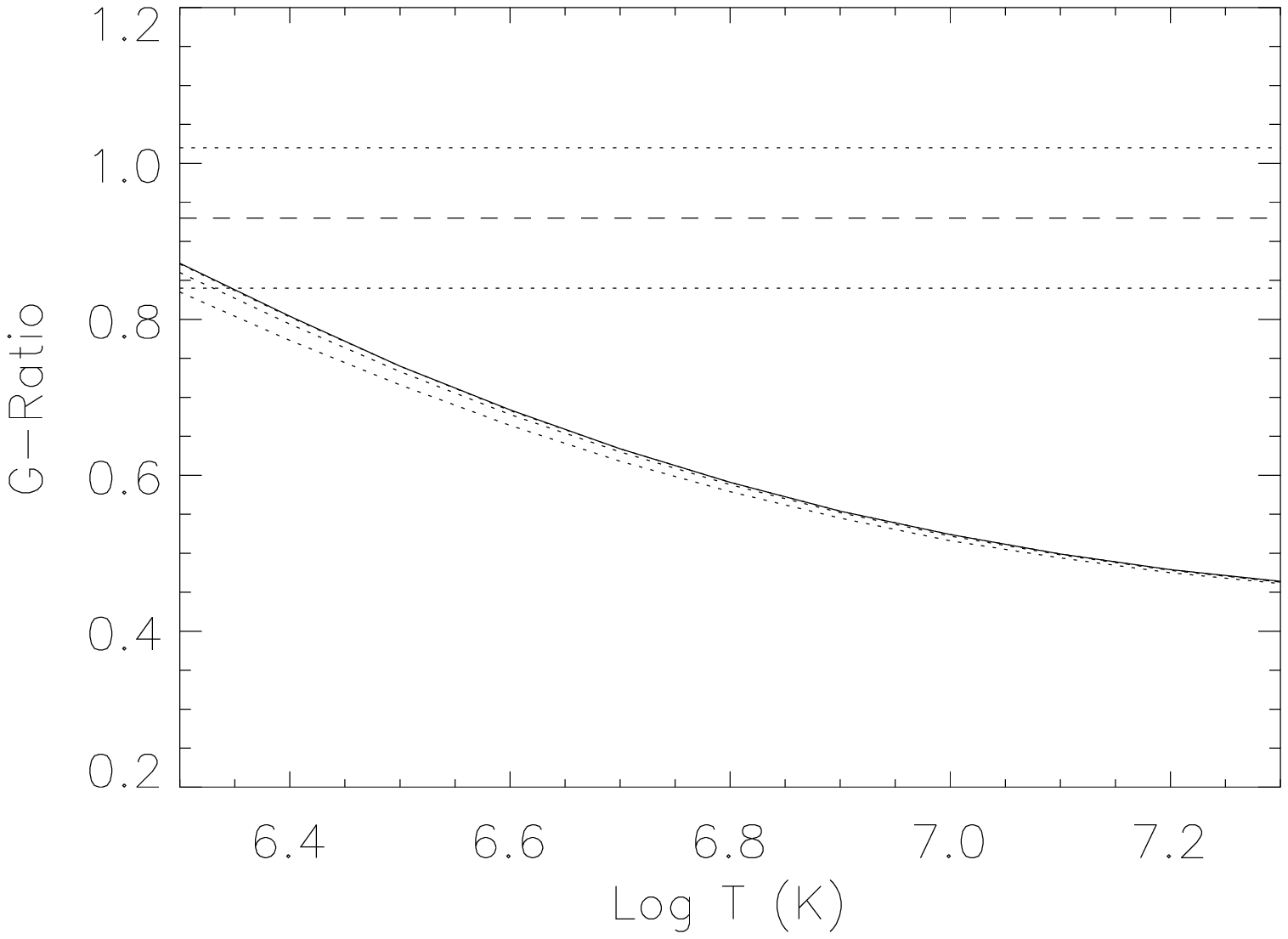}}
\caption{\label{temp_plot}Determination of the plasma temperature from
the $G$-ratio for the fluxes adjusted for line blending (dashed line).
Errors on the measured ratio are represented by dotted lines. APEC models are
given for $10^{10}$, $10^{11}$, $10^{12}$, and $10^{13}$\,cm$^{-3}$, with the
solid curve denoting the $10^{10}$\,cm$^{-3}$ (i.e., low-density) model.}
\end{figure}

The $G$-ratio is less affected by blending. The raw measurements
result in $G=0.97 \pm 0.07$, while the deblended fluxes give $G=0.93
\pm 0.09$, which leads to T$=2.1$ $\pm$ 0.9\,MK
(Fig.~\ref{temp_plot}).

\begin{figure}[!ht]
\resizebox{\hsize}{!}{\includegraphics{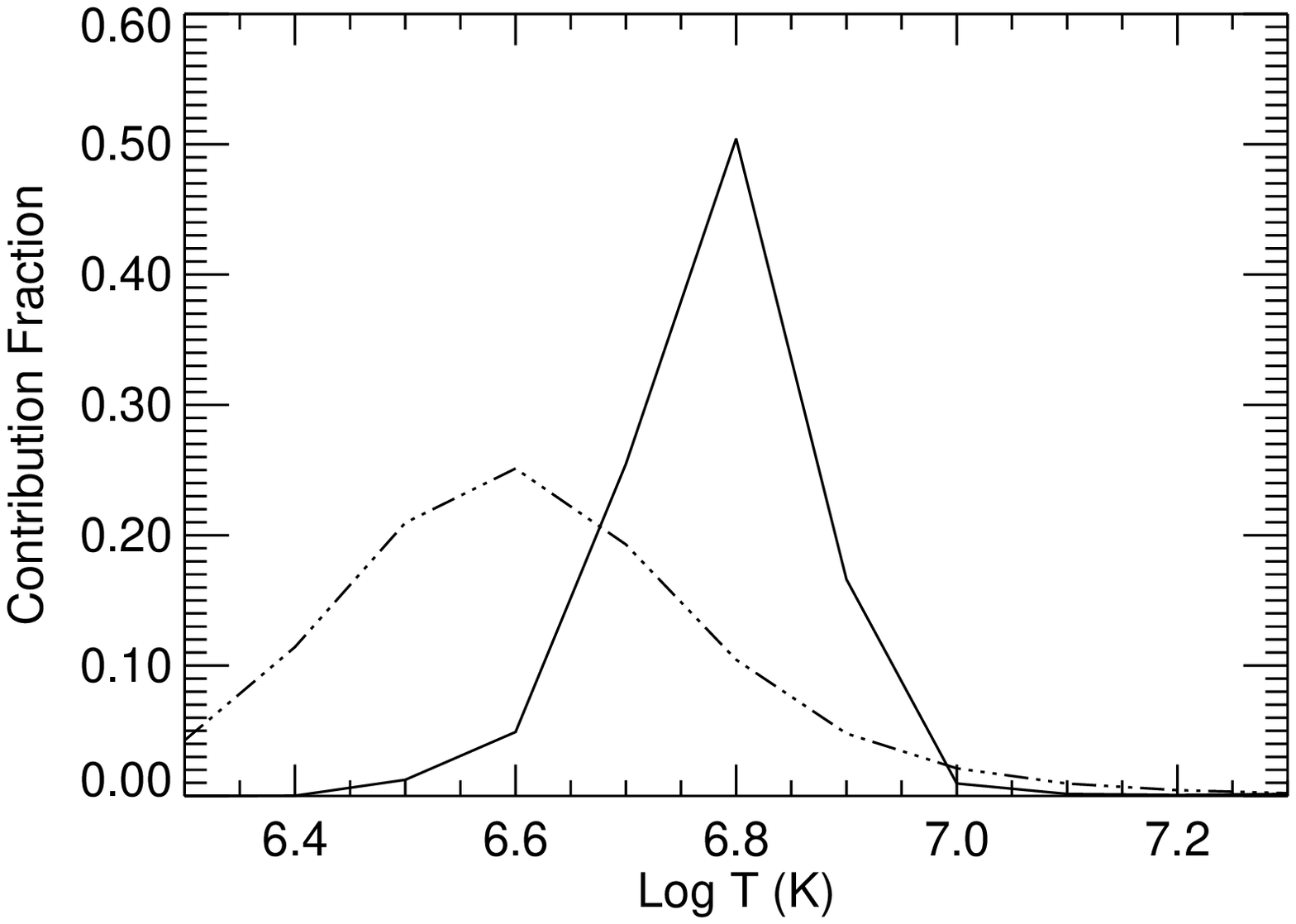}}
\caption{\label{contrib_plot}The fraction of \ion{Ne}{9} resonance
line emission arising from each temperature in the plasma (solid
curve), derived by multiplying the emission measure distribution of
Figure~\ref{emmodel} with the fractional emissivity curve from APEC
(dash-dotted). The intercombination and forbidden lines have virtually
the same dependences and are not shown.}
\end{figure}

The temperature derived from the $G$-ratio is a factor of two lower
than the temperature of maximum emissivity (4\,MK), but probably
consistent given the uncertainties. However, it is inconsistent with
expectations based on the emission measure
distribution. Figure~\ref{contrib_plot} shows that $\sim94$\% of the
line emission in our model comes from temperatures above 4\,MK. This
discrepancy is perhaps most easily interpreted as an abundance effect
in an inhomogeneous plasma, though could also be caused by other
effects.  Standard emission measure analysis is not valid unless the
emitting plasma has uniform abundances; this could easily be incorrect
for active binary systems.

If we allow for this possibility in the case of Capella, we could
conclude that the dominant 6\,MK plasma predicted by the peak in the
emission measure distribution must have a lower than solar
neon-to-iron ratio.  However, if this were the case, in order to
preserve the approximately solar neon-to-iron abundance ratio
indicated by our line-to-continuum analysis (the scaling factor
$A_r=1.03$ for the Ne~IX resonance line), the 2\,MK plasma must have a
much higher neon-to-iron ratio.

\cite{2001ApJ...555L.121Y} have used FUSE to show that the
\ion{Fe}{18} emission formed at 6\,MK is associated predominantly with
the G8 giant, while \cite{2002ApJ...565L..97J}, using HST/STIS, have
found \ion{Fe}{21} emission at $\sim10$\,MK to be predominantly on the
rapidly rotating G1 giant.  These latter observations were concurrent
with some of those included in this analysis.  Earlier measurements
with HST/GHRS had found each star contributing roughly half the
\ion{Fe}{21} emission \citep{1998ApJ...492..767L}.

Our results
suggest that a high neon-to-iron abundance ratio is associated with
the G1 star.  Such a result in the case of the more rapidly rotating
star would not be unprecedented. As noted in \S\ref{usehe}, several
analyses of {\it Chandra} and XMM-Newton spectra of active stars
indicate abundance ratios for neon-to-iron considerably in excess of
the accepted solar value.  \cite{2001ApJ...548L..81D} also noted
similar results arose in earlier analyses of low resolution ASCA
studies. If the G1 Hertzsprung gap giant has $3 \times$ solar
neon-to-iron ratio in its corona and contributes only 25\% at the
emission measure distribution peak (based on the FUSE \ion{Fe}{18}
measurement), then the clump giant would
need to have less than about 40\% of the solar neon-to-iron abundance
ratio. A self-consistent analysis is difficult without more stringent
constraints on the emission measure distribution, especially below
$6$~MK.  However, though the solar FIP effect presents an existing
observational framework for a plausible neon-to-iron ratio
significantly below the solar value, it seems coincidental that
the abundance ratio of the G8 clump giant would conspire to produce a 
global average neon-to-iron abundance ratio which agreed so accurately
with that of the Sun. Furthermore, preliminary analyses of the clump
giants $\gamma$~Tau and $\beta$~Cet do not indicate low neon-to-iron
ratios (Brickhouse \& Dupree 2003, in preparation; Drake et al. 2003,
in preparation, respectively).

While we cannot totally rule out line blending, the required
contamination would be far larger than expected. If the $G$-ratio were
high because of a contaminated $f$ line, the contaminating line would
be the fifth strongest iron line in this region, and therefore, hard
for us to have missed. The discrepancy might also arise from errors in the ionization
balance. The shape of the emission measure distribution, derived
primarily from the ionization balance of iron, is subject to
significant atomic data uncertainties
\citep{Brickhouse:Raymond:95}. One further possible effect concerns
the breakdown of the fundamental assumptions implicit in the coronal
approximation: that the plasma is optically thin and is in ionization
equilibrium. A consistent treatment of the inhomogeneities in the
system (two stars with different coronal structures contributing to
the emission) is needed in order to test these assumptions.  We defer
further study of this complicated issue to future work.

For Capella we have used iron lines observed in HETG and EUVE to
determine the emission measure distribution. For observations of other
stellar coronae, which may have less sampling of the iron ionization
stages or lower signal-to-noise than we have for Capella, the
construction of an emission measure distribution becomes more
complicated. A different approach is suggested \citep[e.g.,][]{abun},
where the emission measure distribution is constructed by use of the
H-like Ly$\alpha$ and He-like resonance lines, including a model
distribution of magnetic loops.

\subsection{Blending as a Function of Plasma Temperature} 

As was noted in \S\ref{usehe}, spectra of stellar coronae which are
dominated by plasma at significantly lower temperatures than that of
Capella show considerably less blending around the Ne~IX features,
because the populations of the blending ions (primarily Fe~XIX) are
small.  It is of interest to examine this more quantitatively in order
to determine the temperature regimes for which Ne~IX can be easily
used.

\begin{figure}[!ht]
\resizebox{\hsize}{!}{{\includegraphics{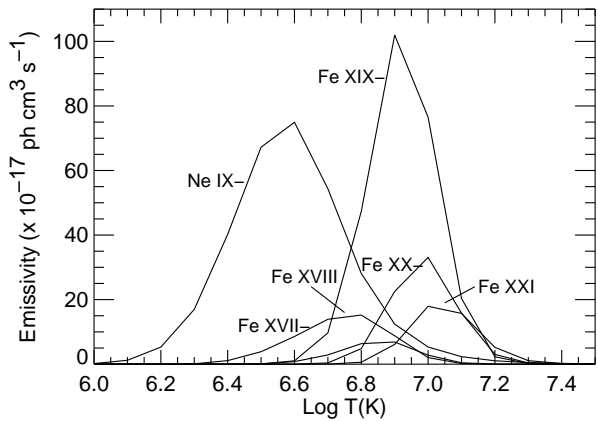}}}
\caption{\label{aped}Total emissivities as functions of temperature for each
ion emitting strong lines in the wavelength range
between 13.35 and 13.85 \AA. Calculations are APEC models
for solar abundances and low density. Table~\ref{feid} gives the
wavelengths for the lines identified in this work.
}
\end{figure}

Figure~\ref{aped} shows the cooling functions for the neon and
blending iron lines.  It is apparent that, in plasmas with
temperatures $\log T$ below $\sim 6.8$, the measurement of \ion{Ne}{9}
will not be severely complicated by blending. In plasmas with higher
temperatures, the blending will be very strong unless neon is
significantly overabundant compared to iron.  Large neon-to-iron, as
well as neon-to-hydrogen, ratios seem common among active stars, such
as in HR~1099 \citep{2001A&A...365L.324B, 2001ApJ...548L..81D}, II~Peg
\citep{2001ApJ...559.1135H}, and AR~Lac \citep{Huenemoerder:03}.
Generally, in all kinds of low-pressure plasmas with high densities
but low temperatures the blending with \ion{Fe}{19} is less severe,
and thus disentangling the \ion{Ne}{9} lines should be fairly
straightforward.

\section{Conclusions}
\label{conc}

The \ion{Ne}{9} triplet is an important tool for estimating plasma
densities, not only for coronal plasmas as in Capella, but for plasmas
in general in the density range between $10^{10}$ and
$10^{13}$\,cm$^{-3}$.  However, we have shown that the
intercombination line of Ne~IX is severely blended in plasma with
temperatures $\log T\ga 6.8$.  This is the case in active coronae as
well as in solar flares, such that deblending of high temperature
lines is also important for solar flare diagnostics.

Through our detailed study of Capella spectra, we have found that APEC
models are sufficiently accurate and complete that all the significant
observed lines in the \ion{Ne}{9} region can be reasonably identified
in the HEG spectra. The laboratory wavelengths of \ion{Fe}{19} from
\cite{2002ApJS..140..589B} provide a significant improvement to the
accuracy over the wavelengths derived from the HULLAC energy levels.
The model fluxes for these lines are also in good agreement with the
observations.  In the APEC models for Capella, the \ion{Ne}{9}
intercombination line is significantly blended. Given the predicted
flux of the blending lines, the $f/i$ ratio is consistent with the low
density limit ($n_e < 2 \times 10^{10}$ cm$^{-3}$).

Interestingly, we find that the temperature-sensitive $G$-ratio is
inconsistent with the emission measure distribution derived from iron
in the sense that it indicates significantly cooler electron
temperatures.  Since the strong peak in the emission measure
distribution at 6~MK is likely produced predominantly by the G8 star
\citep{2001ApJ...555L.121Y}, abundance differences between the two
coronae at first seem the most likely explanation for this apparent
inconsistency.  This would require the neon-to-iron abundance ratio to
be lower in the G8 corona than in the G1 corona.  In contrast, the
neon-to-iron ratio in the G1 corona would have to be higher than solar
values such that the average ratio in observations of the Capella
system as a whole appeared solar.

If correct, this interpretation could have significant implications
for coronal analyses in general, since many coronal sources are binary
systems comprised of two coronae.  Furthermore, if coronae on
individual stars are compositionally inhomogeneous (as is the case on
the Sun), models will have to account for this.  In future work, we
will investigate more detailed models to determine what abundance and
temperature differences are required to explain the Capella $G$-ratio
and emission measure distribution inconsistency. We will also continue
the detailed assessment of blending and atomic data for these and
other spectra, which may shed further light on this problem. 

In the case of Capella spectra obtained by LETGS and RGS, the spectral
resolution is inadequate to derive the same results independently from
this isolated spectral region. We have not yet explored the
possibility of using other stronger, isolated lines from \ion{Fe}{19}
to \ion{Fe}{21} to specify the iron contribution to the blended
\ion{Ne}{9} spectral region, and then perform a more constrained
fit. The potential problem with this approach is that typical
uncertainties in the APEC model line flux ratios ($\sim 20$ to $30$\%)
may be too large to sufficiently constrain the blending spectrum. With
a single test case such as Capella, we cannot explore all the
parameter space in the models, and thus any conclusions concerning the
treatment of blending cannot yet be generalized.  Nevertheless, the
good agreement between the APEC models and the HEG observations is
encouraging.


\acknowledgments

We thank Manuel G\"udel and Marc Audard for support in reducing the
RGS data and Priya Desai for help with the Chandra data reduction. We
also thank Greg Brown for making the pre-publication data from the
LLNL EBIT available to us. We acknowledge support for J.-U. Ness from
NASA LTSA NAG5-3559.  J.~J.~D. and N.~S.~B. were supported by NASA
contract NAS8-39083 to SAO for the CXC and D.~P.~H. was supported by
NASA through Smithsonian Astrophysical Observatory (SAO) contract
SVI-61010 to MIT for the CXC.

\end{document}